\documentclass{article}
\newtheorem{Assumption}{Assumption}
\newtheorem{Definition}{Definition}

\newtheorem{remark}{remark}
\newtheorem{theorem}{theorem}
\newtheorem{proof}{proof}

\usepackage{PRIMEarxiv}
\usepackage{amsmath}
\usepackage{amssymb}
\usepackage{CJKutf8}
\usepackage[utf8]{inputenc} 
\usepackage[T1]{fontenc}    
\usepackage{hyperref}       
\usepackage{url}            
\usepackage{booktabs}       
\usepackage{amsfonts}       
\usepackage{nicefrac}       
\usepackage{microtype}      
\usepackage{lipsum}
\usepackage{fancyhdr}       
\usepackage{graphicx} 
\graphicspath{{media/}}     
\usepackage{algorithm}
\usepackage{algpseudocode}
\usepackage{epstopdf}

\title{Event triggered optimal formation control for nonlinear multi-agent systems under Denial-of-Service attacks

}

\author{
  Zhang Jianqiang\quad\quad   Kaijun Yang \\
  School of Electrical and Control Engineering, Shaanxi University of Science and Technology\\
  Xi'an \quad  China\\
  \texttt{\{Kaijun Yang\}yangkj2020@126.com} \\
}

\begin{document}
\maketitle

\begin{abstract}
This paper investigates the optimal formation control problem of a class of nonlinear multi-agent systems(MASs) under Denial-of-Service(DoS) attacks. We design the optimal formation control law using an event-triggered control scheme to achieve formation objectives under DoS attacks. Critic neural network (NN)-based approach is employed to achieve the optimal control policy under DoS attacks. Event-triggered mechanism is introduced to ensure the saving of control resources.  Additionally,  Lyapunov stability theory is utilized to demonstrate that the local neighborhood formation error exhibits exponential stability and the estimation error of weights are uniformly ultimately bounded. Finally, the effectiveness of the control algorithm is validated through matlab simulations. The results indicate that under DoS attacks, the nonlinear MAS successfully achieves the desired formation for the MAS.
\end{abstract}

\keywords{ multi-agent system\and formation control \and DoS attacks \and optimal control
\and   event-triggering control }

\section{INTRODUCTION}\label{sec1}
Multi-agent formation control addresses the coordination of motion and behavior among multiple agents to achieve and sustain specific  configurations.  The devised control protocols  utilize information exchange among neighboring agents to enable a group of autonomous agents to achieve and maintain a desired geometric formation in their states/outputs. Today, MAS formation control plays a crucial role in various fields such as cooperative monitoring \cite{mcarthur2004design}, aerospace\cite{han2013cooperative}, autonomous underwater robots\cite{chen2020multi}, multi-robot systems\cite{ren2008distributed}, etc.

Nowadays, numerous control techniques have been applied to MASs to achieve the desired formation, including backstepping \cite{zhang2023finite}, fuzzy control\cite{li2021fuzzy}, adaptive sliding mode \cite{fei2020neural}.
In many industrial systems, it is often required for the controlled system to achieve the desired control effect with minimal control effort. Therefore, Optimal formation control of MASs  is proposed  to ensure the stability of the MASs while minimizing performance metrics, thereby achieving the pre-defined formation \cite{zhang2014leader,cai2020fuzzy,zhao2023fuzzy,tang2023adaptive,zhang2017distributed}. To obtain the optimal control strategy, solving the Hamilton-Jacobi-Bellman (HJB) equation becomes critical. Since the HJB is a complex nonlinear  partial differential system, it is very difficult to solve using traditional methods. Adaptive dynamic programming (ADP) method has been widely applied to solve the optimal control problem. Its core idea is to utilize function approximation structures to approximate the solution to HJB. 
References\cite{zhang2014leader,cai2020fuzzy,zhao2023fuzzy} proposed a fuzzy ADP optimal control scheme to address the optimal coordination control problem in MASs. References\cite{tang2023adaptive,zhang2017distributed} achieved optimal control of nonlinear MASs by constructing an actor-critic neural network (NN) framework and updating the neural network weights using gradient descent method.

In network systems, the completion of a task is often threatened by DoS attacks, which can render the target network incapable of providing normal communication. Therefore, effectively mitigating DoS attacks is crucial in MAS formation control. To address the impact of DoS attacks, References \cite{amullen2016model,yang2020observer} investigated secure tracking consensus of nonlinear MASs under continuous and intermittent DoS attacks, designing distributed fixed-time observers to rapidly estimate leader information. References \cite{amullen2016secured,li2022fully} proposed a secure formation control strategy for linear MASs, enabling agents to recover their states under attack. References\cite{dong2020leader,tian2022event,li2024bipartite} investigated the consensus problem of linear MASs under DoS attacks. By setting upper bounds on the DoS attack frequency and attack length rate, they ensured that the system can achieve consensus. Overall, existing literature lacks sufficient research on optimal formation control for MASs in the presence of DoS attacks.

In many practical applications, the communication capability between agents is often limited. High-density communication can lead to significant communication burdens. Therefore, to address this issue, event-triggered control has received widespread attention.The principle of event-triggering mechanisms is based on initiating control signal updates only when specific conditions are met, while maintaining the control signal when the system is stable. References\cite{pham2023adaptive,luo2023event,cai2021fixed} proposed a distributed event-triggered control scheme for nonlinear MASs, addressing the formation tracking problem in such systems.  When addressing the optimal formation control problem, integrating ADP with event-triggered control methods can offer substantial advantages in alleviating the control burden.  Event-triggered ADP method was proposed for MASs to address the trajectory tracking \cite{wang2022event2}, the average consensus \cite{zhao2019event,zhao2019distributed} , and formation tracking problems \cite{dou2022event}.

Inspired by the aforementioned works, this paper explores the event-triggered optimal formation control problem for nonlinear MASs in the presence of DoS attacks.
The main contributions of this paper can be summarized as follows:
\begin{enumerate}
	\item Optimal formation control of  MASs under DoS attacks is the focus of our study. Unlike the existing references \cite{zhao2019event,zhao2019distributed,wang2022event}, which primarily addressed event-triggered and distributed control strategies, we specifically investigate the impact of DoS attacks on formation control for MASs. These attacks can severely disrupt the control performance of the systems. Our proposed algorithm is designed to determine the safe upper bounds of attack length rate and frequency, ensuring that nonlinear MASs can effectively achieve formation control even under such adverse conditions.
	\item  Event-triggered mechanisms for each agent are distributed, operating independently of global information. Compared to traditional ADP methods\cite{zhang2014leader,cai2020fuzzy}, our event-triggered ADP formation control method only requires updating the neural network weights at the event-triggering instants, which effectively saves computational costs. 
	\item By employing Bellman optimality principle, the relationship between the optimal control and the value function is established. We approximate the value function using a critic neural network and solve the coupled HJB equation to derive the optimal control policy and optimal performance index function. Compared to traditional actor-critic dual-network structures\cite{tang2023adaptive,liu2023aperiodically}, this single-network architecture simplifies the design of the optimal formation control strategy.
\end{enumerate}
The remaining parts of this paper are organized as follows. Section \ref{sec2} formulates the problem and introduces necessary background knowledge on graph theory and DoS attacks. Section \ref{sec_3} utilizes optimal control methods and event-triggered control methods to design event-triggered optimal control laws, and rigorously analyzes the stability of the controlled system under DoS attacks, subsequently, the PI algorithm implementation of the optimal control strategy is presented. Section \ref{sec_4} extends the PI algorithm from Section \ref{sec_3}, using neural network methods to approximate the optimal control strategy, and proves the convergence of weight errors and the feasibility of the event-triggering mechanism. In Section \ref{sec_5}, simulations are conducted to demonstrate the effectiveness of the theoretical results. Finally, Section \ref{sec_6} provides concluding remarks.

\section{PROBLEM STATEMENT AND PRELIMINARIES}\label{sec2}
\subsection{Graph theory}
This section will utilize graph theory to illustrate the communication relationships among intelligent agents. Considering a MAS composed of $N$ agents, the communication topology among the agents can be represented as graph $G=\text{ }\left( \nu ,\text{ }\varepsilon ,\text{ }A \right)$, where $\nu$ represents the set of all vertices, $\text{ }\varepsilon$ represents the set of edges, and $A$ is the adjacency matrix. In the communication topology graph, vertices represent the positions of each follower agent, and edges signify the exchange of information between the agents, adjacency matrix $A=\lbrace{{a}_{ij}}\rbrace\in {{\mathbb{R}}^{N\times N}}$. When the communication topology is an undirected graph, ${{a}_{ij}}={{a}_{ji}}$. If $(i,j)\notin \varepsilon$, then ${{a}_{ij}}=0$, and if $(i,j)\in \varepsilon$, then ${{a}_{ij}}=1$. Here, $(i,j)\in \varepsilon$ signifies that agent $i$ can transmit its own state information to agent $j$, The set of neighboring vertices for node $i$ can be represented as ${{\mathcal{N}}_{i}}=\left\{ j\in \nu :\left( i,j \right)\in \varepsilon  \right\}$. In addition, the Laplacian matrix for an undirected communication topology is defined as $L=D-A$, where $D=\operatorname{diag}\{{{d}_{1}},\cdots ,{{d}_{n}}\}$ is the in-degree matrix of the nodes, and ${{d}_{i}}=\sum{_{j=1}^{N}{{a}_{ij}}}$ for $i=1,2,\cdots N$. The connection matrix between leaders and followers is defined as $B=\operatorname{diag}\{{{b}_{1}},\cdots ,{{b}_{n}}\}$. When follower agent $i$ can receive information from leader agent, ${{b}_{i}}>0$, otherwise,  ${{b}_{i}}=0$. 
\subsection{ Problem formulation}
Consider a nonlinear MAS composed of one leader and $N$ followers, the dynamics model for the 
$i$-th follower can be described as
\begin{eqnarray}\label{xuhao1}
	{{\dot{x}}_{i}}(t)=A{{x}_{i}}(t)+B{{u}_{i}}(t)+f_{i}({{x}_{i}}(t)),
\end{eqnarray}
where $ x_{i}\in \mathbb{R}^n $  and $ u_{i}\in \mathbb{R}^n $ are the state and control input of agent $i$, respectively, $f_{i}(\cdot )$ is an unknown nonlinear function. $A$ and $B$ are constant matrices with appropriate dimension. 
The dynamics of the leader can be considered as
\begin{eqnarray}\label{xuhao2}
	{{\dot{x}}_{0}}(t)=f_{0}({{x}_{0}}(t)).
\end{eqnarray}
The vector $ x_{0}\in \mathbb{R}^n $ is the state  of the leader, respectively.  
The local neighborhood formation error $e_i$ for the $i$-th agent is defined as
\begin{eqnarray}\label{xuhao3}
	\begin{aligned}
		e_{i} =&\sum_{j\in N_{i} } a_{ij} (x_{i}-\iota _{i} -x_{j}+\iota _{j}   )+b_{i} (x_{i}-x_{0} -\iota _{i}  ).
	\end{aligned}
\end{eqnarray}
The parameter $\iota _{i}$ represents the relative position between the $i$-th follower and the leader during the formation process.
Further, according to \eqref{xuhao1}, \eqref{xuhao2} and \eqref{xuhao3}, the dynamic description of the formation error can be obtained as
\begin{equation}\label{xuhao4}
	\begin{aligned}
		{{{\dot{e}}}_{i}}&= \sum\limits_{j\in {{N }_{i}}}{{{a}_{ij}}(A{{x}_{i}}+B{{u}_{i}}+f_{i}({{x}_{i}}))-A{{x}_{j}}-B{{u}_{j}}-f_{j}({{x}_{j}}))}+{{b}_{i}}(A{{x}_{i}}+B{{u}_{i}}+f_{i}({{x}_{i}})-f_{0}({{x}_{0}})) \\ 
		& =A{{e}_{i}}+B({{d}_{i}}+{{b}_{i}}){{u}_{i}}-B\sum\limits_{j\in {{N }_{i}}}{{a}_{ij}}{{u}_{j}} + \sum\limits_{j\in {{N }_{i}}}{{{a}_{ij}}(f_{i}({{x}_{i}})-f_{j}({{x}_{j}}))}+{{b}_{i}}(f_{i}({{x}_{i}})-f_{0}({{x}_{0}})) \\ 
		&\;\;\; \text{  }+A\Big[ \sum\limits_{j\in {{N}_{i}}}{{{a}_{ij}}({{\iota}_{i}}-{{\iota}_{j}})+{{b}_{i}}{{\iota}_{i}}} \Big] \\ 
		& =A{{e}_{i}}+{{B}_{1}}{{u}_{i}}-B\sum\limits_{j\in {{N }_{i}}}{{a}_{ij}}{{u}_{j}}+{{F}_{i}}+C, 
	\end{aligned}
\end{equation}
where ${B}_{1}=B({{d}_{i}}+{{b}_{i}})$, ${{F}_{i}}=\sum\limits_{j\in {{N}_{i}}}{{{a}_{ij}}(f_{i}({{x}_{i}})-f_{j}({{x}_{j}}))}+{{b}_{i}}(f_{i}({{x}_{i}})-f_{0}({{x}_{0}}))$, $C=A\Big[ \sum\limits_{j\in {{N}_{i}}}{{{a}_{ij}}({{\iota}_{i}}-{{\iota}_{j}})+{{b}_{i}}{{\iota}_{i}}} \Big]$.\\
\subsection{DoS attack model}

To study the stability of the nonlinear MAS when the communication channel is subjected to a DoS attack, the following DoS attack model is employed.  Let $\{{{h}_{n}}\}$ be the time sequence when DoS attacks occur. Define ${{D}_{n}}=[{{h}_{n}},{{h}_{n+1}})$ as the time interval of the $n$-th DoS attack with duration $\tau (n)$ during which the communication between the controller and the system's actuator is disrupted. Based on the definitions of ${{h}_{n}}$, ${{D}_{n}}$ and ${{\tau }_{n}}$,  the time interval $[0,t]$ can be divided into the following cases:
\begin{enumerate}
	\item The function $\Theta (t)$ represents the time interval $[0, t)$ excluding the time under network attacks and is defined as
	\begin{eqnarray}\label{xuhao16}
		\begin{aligned}
			\Theta (t): =[0,t]\backslash {{D}_{n}}.
		\end{aligned}
	\end{eqnarray}
	\item The function $\Omega (t)$ represents the time interval $[0, t)$ under network attacks and is defined as
	\begin{eqnarray}\label{xuhao17}
		\Omega (t): ={{D}_{n}}\bigcap [0,t].
	\end{eqnarray}
\end{enumerate}
After the above analysis, when the MAS is not subjected to DoS attacks, the aforementioned event-triggered control signal is utilized. When the MAS is under DoS attacks, the communication channels between agents will be interrupted. In this case, agent $i$ will not communicate with any of its neighbors, i.e. ${{u}_{i}}(t)=0$. To facilitate the analysis of the impact of attacks on MAS performance, the following definition is provided.
\begin{Definition}\label{Def1}\cite{dong2020leader}
	Let $N(0, t)$ denote the number of DoS attacks within the time interval $[0, t)$. The attack frequency is defined as follows
	\begin{eqnarray}\label{xuhaog1}
		\begin{aligned}
			F(0,t)=\frac{N(0,t)}{t}.
		\end{aligned}
	\end{eqnarray}
\end{Definition}
\begin{Definition}\label{Def2}\cite{dong2020leader}
	For any $t > 0$, let $\Omega(t)$ denote the total duration of attacks within the time interval $[0, t)$. The attack length rate is defined as follows
	\begin{eqnarray}\label{xuhaog2}
		\begin{aligned}
			T(0,t)=\frac{\Omega (t)}{t}.
		\end{aligned}
	\end{eqnarray}
\end{Definition}
The control objective of this paper is to design an optimal control law such that the nonlinear MAS described above can form the desired formation even in the presence of network attacks while maintaining consistency with the leader's behavior. Therefore, the performance index function for the agent $i$ is defined as
\begin{eqnarray}\label{xuhao6}
	\begin{aligned}
		{{J}_{i}}=\int_{0}^{\infty }{(e_{i}^{T}{{Q}_{ii}}{{e}_{i}}+u_{i}^{T}{{R}_{ii}}{{u}_{i}}+\sum\limits_{j\in {{N }_{i}}}{u_{j}^{T}{{R}_{ij}}{{u}_{j}}})dt},
	\end{aligned}
\end{eqnarray}
where ${{Q}_{ii}}>0$, ${{R}_{ii}}>0$, and ${{R}_{ij}}\ge 0$ are constant matrices.

\begin{remark}  
	The formation conditions require that the duration and frequency of DoS attacks must be below a specified threshold. Once the length rate or frequency of a DoS attack exceeds this threshold, the control strategy becomes ineffective. This means that as long as the DoS attack lasts long enough, it can lead to the formation control failure and result in the control strategy fails.
\end{remark}
\section{Event-triggered Optimal Controller Design}\label{sec_3}
This section proposes the optimal control strategy based on event-triggering mechanisms. It also provides sufficient conditions for achieving optimal secure formation under DoS attacks. Additionally, the process of solving coupled HJB equations using the Policy Iteration (PI) algorithm is presented.

Under control strategies $u_i$ and $u_j$, the value function $V_i(e_i)$ of node $i$ is defined as
\begin{equation}\label{xuhao7}
	\begin{aligned}
		{{V}_{i}}({{e}_{i}})=\int_{0}^{\infty }{(e_{i}^{T}{{Q}_{ii}}{{e}_{i}}+u_{i}^{T}{{R}_{ii}}{{u}_{i}}+\sum\limits_{j\in {{N }_{i}}}{u_{j}^{T}{{R}_{ij}}{{u}_{j}}})dt}.
	\end{aligned}
\end{equation}
To find the control law $u_i^*$ that minimizes the performance index $J_i$, the following Hamiltonian function is defined
\begin{equation}\label{xuhao8}
	\begin{aligned}
		{{H}_{i}}({{e}_{i}},\nabla {{V}_{i}},{{u}_{i}},{{u}_{(i)}})=e_{i}^{T}{{Q}_{ii}}{{e}_{i}}+\sum\limits_{j}{u_{j}^{T}{{R}_{ij}}{{u}_{j}}}+\nabla V_{i}^{T}({{\dot{e}}_{i}}),
	\end{aligned}
\end{equation}
where $\nabla {{V}_{i}}=\frac{\partial {{V}_{i}}}{\partial {{e}_{i}}}$.\\
The optimal value function $V_{i}^{*}({{e}_{i}})$ satisfies the following function
\begin{eqnarray}\label{xuhao9}
	\begin{aligned}
		\underset{{{u}_{i}}}{\mathop{\min }}\,H({{e}_{i}},\nabla V_{i}^{*},{{u}_{i}},{{u}_{(j)}})=0.
	\end{aligned}
\end{eqnarray}
Then the optimal control law can be expressed as
\begin{eqnarray}\label{xuhao10}
	\begin{aligned}
		u_{i}^{*}=-\frac{1}{2}({{d}_{i}}+{{b}_{i}}){{R}_{ii}}^{-1}{{B}^{T}}\nabla V_{i}^{*}.
	\end{aligned}
\end{eqnarray}
The corresponding local optimal Hamilton–Jacobi–Bellman for the $i$-th agent is
\begin{equation}\label{xuhao11}
	\begin{aligned}
		0&={{H}_{i}}({{e}_{i}},\nabla V_{i}^{*},u_{i}^{*},u_{(j)}^{*}) \\ 
		& ={{e}_{i}}^{T}{{Q}_{ii}}{{e}_{i}}+{{u}_{i}}^{*T}{{R}_{ii}}{{u}_{i}}^{*}+\sum\limits_{j\in {{N}_{i}}}{{{u}_{j}}^{*T}{{R}_{ij}}{{u}_{j}}^{*}}\\
		&\;\;\;\; +\nabla V_{i}^{*T}\Big(A{{e}_{i}}+{{B}_{1}}{{u}_{i}^{*}}-B\sum\limits_{j\in {{N }_{i}}}{{a}_{ij}}{{u}_{j}^{*}}+{{F}_{i}}+C\Big).
	\end{aligned}
\end{equation}

In order to reduce the consumption of communication resources between agents while maintaining a good formation effect in the MAS, a distributed event-triggered control method is employed for control. The control signal can be expressed as
\begin{eqnarray}\label{xuhao5}
	{{u}_{i}^{*}}(t)={{u}_{i}^{*}}(t_{k}^{i}),\text{  }t\in [t_{k}^{i},t_{k+1}^{i}),
\end{eqnarray}
where $t_{k}^{i}$ represents the triggering time sequence, and the event triggering time is defined as
\begin{eqnarray}\label{xuhaogg5}
	t_{k+1}^{i}=\inf \{t>t_{k}^{i}|{{g}_{i}}(t)>0\},
\end{eqnarray}
where  ${{g}_{i}}(t)>0$ as the event-triggering function will be presented later.\\
Additionally, the measurement error is defined as 
\begin{eqnarray}\label{xuhaogg6}
	{{\delta }_{i}}(t)={{e}_{i}}(t_{k}^{i})-{{e}_{i}}(t).
\end{eqnarray}
Combining the optimal control strategy \eqref{xuhao10} and event-triggered control strategy \eqref{xuhao5}, the control protocol for a nonlinear MAS under DoS attacks is given as follows
\begin{eqnarray}\label{xuhao12}
	\begin{aligned}
		u_{i}^{*}(t_{k}^{i})=-\frac{1}{2}({{d}_{i}}+{{b}_{i}}){{R}_{ii}}^{-1}{{B}^{T}}\nabla V_{i}^{*}(t_{k}^{i}).
	\end{aligned}
\end{eqnarray}
Then the error system \eqref{xuhao4} can be rewritten as
\begin{equation}\label{xuhao13}
	{{\dot{e}}_{i}}(t)=\left\{ \begin{aligned}
		& A{{e}_{i}}(t)+{{B}_{1}}u_{i}^{*}(t_{k}^{i})-B\sum\limits_{j\in {{N }_{i}}}{{a}_{ij}}{{u}_{j}^{*}}(t)+{{F}_{i}}(t)+C,\text{ }t\in [{{h}_{k-1}}+{{\tau }_{k-1}},{{h}_{k}}), \\ 
		& A{{e}_{i}}(t)+{{F}_{i}}(t)+C,\text{ }t\in [{{h}_{k}},{{h}_{k}}+{{\tau }_{k}}). \\ 
	\end{aligned} \right.
\end{equation}
To achieve optimal formation control of nonlinear MAS, the following is the description of the assumptions that will be used in this paper.
\begin{Assumption}\label{Assu1}\cite{yang2020online}
	There exists a constant $\gamma > 0$ such that the non-linear vector-valued function $f(\cdot)$ satisfies the following Lipschitz condition
	\begin{eqnarray}\label{xuhaogg2}
		\begin{aligned}
			\Vert f(\lambda )-f(\mu )\Vert\text{ }\leq \gamma \Vert \lambda -\mu \Vert,\quad \forall \lambda ,\mu \in {{\mathbb{R}}^{n}}.
		\end{aligned}
	\end{eqnarray}
\end{Assumption}
\begin{Assumption}\label{Assu2}\cite{vamvoudakis2014event}
	The event-triggered sampled controller is Lipschitz continuous. That is, there exists a constant $M$ such that 
	\begin{eqnarray}\label{xuhaog6}
		\begin{aligned}
			\Vert u_{i}^{*}-u_{i}^{*}(t_{k}^{i})\Vert \text{ }\le M\Vert {{\delta }_{i}}\Vert.				
		\end{aligned}
	\end{eqnarray}
\end{Assumption}
\subsection{Stability Analysis}
In this subsection, stability of the nonlinear MAS will be proven through the following theorem.
\begin{theorem}\label{thm1}
	Consider the nonlinear MAS \eqref{xuhao1} and \eqref{xuhao2}. Suppose that Assumptions \ref{Assu1} - \ref{Assu2} hold, by designing the event-triggering condition $g(t)=$
	\begin{equation}\label{xuhao015}
		\begin{aligned}
			{\Vert {{\delta }_{i}} \Vert}^{2} - \frac{\sum\limits_{j\in {{N}_{i}}}{{{\lambda }_{\min }}({{R}_{ij}}){{\Vert u_{j}^{*}(t) \Vert}^{2}}+{{\lambda }_{\min }}({{R}_{ii}}){{\Vert u_{i}^{*}(t_{k}^{i}) \Vert}^{2}}}}{\lambda_{\max }^{2}({{R}_{ii}}){{M}^{2}}}.
		\end{aligned}
	\end{equation} 
	If the system satisfies the DoS attack frequency $F(0,t)\le \frac{{{k}^{*}}}{\ln ({\zeta{C}_{4}})}$ and the DoS attack length rate $T<\frac{{{C}_{1}}-{{k}^{*}}}{{{C}_{1}}+{{C}_{2}}}$, $\zeta$, ${k}^{*}$, ${C}_{2}$ and ${C}_{4}$ are positive constants to be given later, then the formation error of the nonlinear MAS exponentially converges to zero in the presence of DoS attacks.
\end{theorem}
\begin{proof}
	In the absence of DoS attacks, considering the Lyapunov candidate function as 
	\begin{eqnarray}\label{xuhaogg1}
		\begin{aligned}
			V_{1}(t)=\frac{1}{2}\sum\limits_{i=1}^{N}{{{e}_{i}^{T}}(t){{Q}_{ii}}{{e}_{i}}(t)}.
		\end{aligned}
	\end{eqnarray}
	According to equation \eqref{xuhao11}, taking derivative of $V_{1}$ along the system \eqref{xuhao13} gives
	\begin{equation}\label{xuhao18}
		\begin{aligned}
			\dot{V}_{1}(t)& =\sum\limits_{i=1}^{N}{\nabla V_{i}^{*T}({{{\dot{e}}}_{i}})} \\ 
			& =\sum\limits_{i=1}^{N}\nabla V_{i}^{*T}(A{{e}_{i}}(t)+{{B}_{1}}u_{i}^{*}(t)-B\sum\limits_{j\in {{N }_{i}}}{{a}_{ij}}{{u}_{j}^{*}}(t)+{{F}_{i}}(t)+C)\\
			&\;\;\;\; +\sum\limits_{i=1}^{N}{\nabla V_{i}^{*T}({{B}_{1}}u_{i}^{*}(t_{k}^{i})-{{B}_{1}}u_{i}^{*})} \\ 
			& =\sum\limits_{i=1}^{N}\bigg[ -e_{i}^{T}{{Q}_{ii}}{{e}_{i}}-u_{i}^{*T}{{R}_{ii}}u_{i}^{*}-\sum\limits_{j\in {{N}_{i}}}{u_{j}^{*T}{{R}_{ij}}u_{j}^{*}} \bigg]\\
			&\;\;\;\; +\sum\limits_{i=1}^{N}{\nabla V_{i}^{*T}({{B}_{1}}u_{i}^{*}(t_{k}^{i})-{{B}_{1}}u_{i}^{*})},
		\end{aligned}
	\end{equation}
	According to the control protocol \eqref{xuhao10}, it can be inferred that
	\begin{eqnarray}\label{xuhao19}
		\begin{aligned}
			&\sum\limits_{i=1}^{N}{\nabla V_{i}^{*T}({{B}_{1}}u_{i}^{*}(t_{k}^{i})-{{B}_{1}}u_{i}^{*})}\\
			&=-2\sum\limits_{i=1}^{N}{u{{_{i}^{*}}^{T}}{{R}_{ii}}(u_{i}^{*}(t_{k}^{i})-u_{i}^{*})} \\ 
			&=-2\sum\limits_{i=1}^{N}{u{{_{i}^{*}}^{T}}{{R}_{ii}}u_{i}^{*}(t_{k}^{i})}+2\sum\limits_{i=1}^{N}{u{{_{i}^{*}}^{T}}{{R}_{ii}}u_{i}^{*}}.
		\end{aligned}
	\end{eqnarray}
	Furthermore, substituting equation \eqref{xuhao19} into equation \eqref{xuhao18} yields
	\begin{equation}\label{xuhao20}
		\begin{aligned}
			\dot{V}_{1}(t)&=\sum\limits_{i=1}^{N}-e_{i}^{T}{{Q}_{ii}}{{e}_{i}}+u_{i}^{*T}{{R}_{ii}}u_{i}^{*}-\sum\limits_{j\in {{N}_{i}}}u_{j}^{*T}{{R}_{ij}}u_{j}^{*} -2u{{_{i}^{*}}^{T}}{{R}_{ii}}u_{i}^{*}(t_{k}^{i}) \\ 
			& =\sum\limits_{i=1}^{N}-e_{i}^{T}{{Q}_{ii}}{{e}_{i}}+u_{i}^{*T}{{R}_{ii}}u_{i}^{*}-\sum\limits_{j\in {{N}_{i}}}u_{j}^{*T}{{R}_{ij}}u_{j}^{*} -2u{{_{i}^{*}}^{T}}{{R}_{ii}}u_{i}^{*}(t_{k}^{i})\\
			&\;\;\;\;+u_{i}^{*T}(t_{k}^{i}){{R}_{ii}}u_{i}^{*}(t_{k}^{i})-u_{i}^{*T}(t_{k}^{i}){{R}_{ii}}u_{i}^{*}(t_{k}^{i}) \\ 
			& =\sum\limits_{i=1}^{N}-e_{i}^{T}{{Q}_{ii}}{{e}_{i}}-\sum\limits_{j\in {{N}_{i}}}u_{j}^{*T}{{R}_{ij}}u_{j}^{*} +{{\left\| {{R}_{ii}}(u_{i}^{*}-u_{i}^{*}(t_{k}^{i})) \right\|}^{2}}-u_{i}^{*T}(t_{k}^{i}){{R}_{ii}}u_{i}^{*}(t_{k}^{i}) \\ 
			& \le \sum\limits_{i=1}^{N}-{{\lambda }_{\min }}({{Q}_{ii}}){{\Vert {{e}_{i}} \Vert}^{2}}-\sum\limits_{j\in {{N}_{i}}}{{\lambda }_{\min }}({{R}_{ij}}){{\Vert u_{j}^{*} \Vert}^{2}} +\lambda _{\max }^{2}({{R}_{ii}}){{M}^{2}}{{\Vert {{\delta }_{i}}\Vert}^{2}}\\
			&\;\;\;\;-{{\lambda }_{\min }}({{R}_{ii}}){{\left\| u_{i}^{*}(t_{k}^{i}) \right\|}^{2}}.
		\end{aligned}
	\end{equation}
	Derived from the event-triggering conditions \eqref{xuhao015}, it can be concluded that 
	\begin{equation}
		{{\Vert {{\delta }_{i}} \Vert}^{2}}\le \frac{\sum\limits_{j\in {{N}_{i}}}{{{\lambda }_{\min }}({{R}_{ij}}){{\Vert u_{j}^{*} \Vert}^{2}}+{{\lambda }_{\min }}({{R}_{ii}}){{\Vert u_{i}^{*}(t_{k}^{i}) \Vert}^{2}}}}{\lambda _{\max }^{2}({{R}_{ii}}){{M}^{2}}}. 
	\end{equation}
	Therefore,   \eqref{xuhao20} can be further estimated as
	\begin{eqnarray}\label{xuhao21}
		\begin{aligned}
			\dot{V}_{1}(t)&\le \sum\limits_{i=1}^{N}{-{{\lambda }_{\min }}({{Q}_{ii}}){{\left\| {{e}_{i}} \right\|}^{2}}} \\ 
			& \le -{{C}_{1}}V_{1}(t),
		\end{aligned}
	\end{eqnarray}
	where ${{C}_{1}}={{\lambda }_{\min }}({{Q}_{ii}})$.\\
	When the system is under DoS attacks, given $P > 0$, consider the following Lyapunov candidate function
	\begin{eqnarray}\label{xuhao22}
		\begin{aligned}
			V_{2}(t)=\frac{1}{2}\sum\limits_{i=1}^{N}{{{e}_{i}^{T}}(t)P{{e}_{i}}(t)}.
		\end{aligned}
	\end{eqnarray}
	Taking its derivative yields
	\begin{equation}\label{xuhao23}
		\begin{aligned}
			\dot{V}_{2}(t)&=\sum\limits_{i=1}^{N}{{{e}_{i}}^{T}(t)P{{{\dot{e}}}_{i}}(t)}+\sum\limits_{i=1}^{N}{{{{\dot{e}}}_{i}}^{T}(t)P{{e}_{i}}(t)} \\ 
			& =\sum\limits_{i=1}^{N}{{{e}_{i}}^{T}(t)PA{{e}_{i}}(t)}+\sum\limits_{i=1}^{N}{{{e}_{i}}^{T}(t)P\bigg[F_{i}(t)+C \bigg]} \\ 
			&\;\;\; \text{ }+\sum\limits_{i=1}^{N}{{{e}_{i}}^{T}(t){{A}^{T}}P{{e}_{i}}(t)}+{{\sum\limits_{i=1}^{N}{\bigg[ F_{i}(t)+C \bigg]}}^{T}}P{{e}_{i}}(t).
		\end{aligned}
	\end{equation}
	According to Assumption \ref{Assu1}, we have
	\begin{eqnarray}\label{xuhao24}
		\begin{aligned}
			\left\| f_{i}({{x}_{i}}(t))-f_{j}({{x}_{j}}(t)) \right\|\le \lambda \left\| ({{x}_{i}}(t)-{{x}_{j}}(t)) \right\|,
		\end{aligned}
	\end{eqnarray}
	\begin{eqnarray}\label{xuhao25}
		\begin{aligned}
			\left\| f_{i}({{x}_{i}}(t))-f_{0}({{x}_{0}}(t)) \right\|\le \lambda \left\| ({{x}_{i}}(t)-{{x}_{0}}(t)) \right\|.
		\end{aligned}
	\end{eqnarray}
	Thus, $F_{i}(t) \le \lambda{{e}_{i}}(t)$. Furthermore, we get
	\begin{equation}\label{xuhao28}
		\begin{aligned}
			\dot{V}_{2}(t)&\le \sum\limits_{i=1}^{N}{{e}_{i}}^{T}(t)(PA+{{A}^{T}}P){{e}_{i}}(t)+2\lambda {{e}_{i}}^{T}(t)P{{e}_{i}}(t)+{{e}_{i}}^{T}(t)PC+{{C}^{T}}P{{e}_{i}}(t) \\ 
			& \le \sum\limits_{i=1}^{N}{{e}_{i}}^{T}(t)(PA+{{A}^{T}}P){{e}_{i}}(t)+2\lambda {{e}_{i}}^{T}(t)P{{e}_{i}}(t) +2{{e}_{i}}^{T}(t)P{{e}_{i}}(t)+2{{C}^{T}}PC \\ 
			& \le {{\lambda }_{\max }}(PA+AP)\sum\limits_{i=1}^{N}{{{\left\| {{e}_{i}}(t) \right\|}^{2}}}+2(\lambda +1)\sum\limits_{i=1}^{N}{{{e}_{i}}^{T}(t)P{{e}_{i}}(t)+2{{C}^{T}}PC} \\ 
			& \le \left[ \frac{{{\lambda }_{\max }}(PA+AP)}{{{\lambda }_{\min }}(P)}+2(\lambda +1) \right]V_{2}(t)+2{{C}^{T}}PC \\ 
			& ={{C}_{2}}V_{2}(t)+{{C}_{3}},
		\end{aligned}
	\end{equation}
	where ${{C}_{2}}=\frac{{{\lambda }_{\max }}(PA+AP)}{{{\lambda }_{\min }}(P)}+2(\lambda +1)$, ${{C}_{3}}=2{{C}^{T}}PC$. Multiplying both sides of the inequality \eqref{xuhao28} by ${{e}^{-{{C}_{2}}t}}$, we get
	\begin{eqnarray}\label{xuhao29}
		\begin{aligned}
			\dot{V}_{2}(t){{e}^{-{{C}_{2}}t}}-{{C}_{2}}{{e}^{-{{C}_{2}}t}}V_{2}(t) &\le {{C}_{3}}{{e}^{-{{C}_{2}}t}},\\
			\frac{d}{dt}(V_{2}(t){{e}^{-{{C}_{2}}t}}) &\le {{C}_{3}}{{e}^{-{{C}_{2}}t}},\\
			V_{2}(t){{e}^{-{{C}_{2}}t}}-V_{2}(0) &\le -\frac{{{C}_{3}}}{{{C}_{2}}}{{e}^{-{{C}_{2}}t}}+{{C}_{3}}.
		\end{aligned}
	\end{eqnarray}
	Furthermore, transposing terms and multiplying both sides of the inequality by ${{e}^{{{C}_{2}}t}}$, we obtain
	\begin{eqnarray}\label{xuhao30}
		\begin{aligned}
			V_{2}(t) &\le -\frac{{{C}_{3}}}{{{C}_{2}}}+(V(0)+{{C}_{3}}){{e}^{{{C}_{2}}t}} \\ 
			& \le (V_{2}(0)+{{C}_{3}}){{e}^{{{C}_{2}}t}} \\ 
			& ={{C}_{4}}{{e}^{{{C}_{2}}t}},
		\end{aligned}
	\end{eqnarray}
	where ${{C}_{4}}=V_{2}(0)+{{C}_{3}}$.\\
	Let the time interval without network attacks for the $k$-th time be $[{{h}_{k-1}}+{{\tau }_{k-1}},{{h}_{k}})$, and the time interval with network attacks for the $k$-th time be $[{{h}_{k}},{{h}_{k}}+{{\tau }_{k}})$. Let that $V(t) = V_{1}(t)$ be applicable for $t \in [{{h}_{k-1}}+{{\tau }_{k-1}},{{h}_{k}})$ and $V(t) = V_{2}(t)$ be applicable for $t \in [{{h}_{k}},{{h}_{k}}+{{\tau }_{k}})$ , According to  \eqref{xuhao21} and  \eqref{xuhao30}, we obtain
	\begin{eqnarray}\label{xuhao31}
		\begin{aligned}
			V(t)\le \left\{ \begin{aligned}
				& {{e}^{-{{C}_{1}}(t-{{h}_{k-1}}-{{\tau }_{k-1}})}}V_{1}({{h}_{k-1}}+{{\tau }_{k-1}}), t \in [{{h}_{k-1}}+{{\tau }_{k-1}},{{h}_{k}}), \\ 
				& {{C}_{4}}{{e}^{{{C}_{2}}(t-{{h}_{k}})}}V_{2}({{h}_{k}}),\;\; t \in [{{h}_{k}},{{h}_{k}}+{{\tau }_{k}}). \\ 
			\end{aligned} \right.
		\end{aligned}
	\end{eqnarray}
	When $t\in [{{h}_{k-1}}+{{\tau }_{k-1}},{{h}_{k}})$, let $\zeta = \max \big\{ \lambda_{\max}(Q_{ii}) /\lambda_{\min}(P), \lambda_{\max}(P) /\lambda_{\min}(Q_{ii}) \big\}$, it can be obtained that
	\begin{eqnarray}
		\begin{aligned}
			V(t)&\le {{e}^{-{{C}_{1}}(t-{{h}_{k-1}}-{{\tau }_{k-1}})}}V_{1}({{h}_{k-1}}+{{\tau }_{k-1}}) \\ 
			& \le {\zeta{C}_{4}}{{e}^{-{{C}_{1}}(t-{{h}_{k-1}}-{{\tau }_{k-1}})}}{{e}^{{{C}_{2}}(t-{{h}_{k-1}})}}V_{2}({{h}_{k-1}}).
		\end{aligned}
	\end{eqnarray}
	According to  \eqref{xuhaogg1}, \eqref{xuhao22} and \eqref{xuhao31}, we obtain 
	\begin{eqnarray}
		\begin{aligned}
			&V_{1}({{h}_{k-1}}+{{\tau }_{k-1}}) \le \zeta V_{2}({{h}_{k-1}^{-}}+{{\tau }_{k-1}^{-}}), \\ 
			&V_{2}({{h}_{k-1}^{-}}+{{\tau }_{k-1}^{-}}) \le {C}_{4}{{e}^{{{C}_{2}}(t-{{h}_{k-1}})}}V_{2}({{h}_{k-1}}),\\
			&V_{2}({{h}_{k-1}}) \le \zeta V_{1}({{h}_{k-1}^{-}}),\\
			&V_{1}({{h}_{k-1}^{-}}) \le {{e}^{-{{C}_{1}}(t-{{h}_{k-1}}-{{\tau }_{k-1}})}}V_{1}({{h}_{k-2}}+{{\tau }_{k-2}}).
		\end{aligned}
	\end{eqnarray}
	In this manner, proceeding iteratively, we can obtain
	\begin{eqnarray}\label{xuhao32}
		\begin{aligned}
			V(t)&\le {\zeta}^{k}{{C}_{4}}^{\frac{k}{2} }{{e}^{-{{C}_{1}}\left| \Theta (t) \right|}}{{e}^{{{C}_{2}}\left| \Omega (t) \right|}}V_{1}(0).
		\end{aligned}
	\end{eqnarray}
	Similarly, when $t\in [{{h}_{k}},{{h}_{k}}+{{\tau }_{k}})$,  we have
	\begin{equation}\label{xuhao33}
		\begin{aligned}
			V(t)&\le {{e}^{{{C}_{2}}(t-{{h}_{k}})}}V_{2}({{h}_{k}}) \\ 
			& \le \zeta{{C}_{4}}{{e}^{{{C}_{2}}(t-{{h}_{k}})}}{{e}^{-{{C}_{1}}(t-{{h}_{k-1}}-{{\tau }_{k-1}})}}V_{1}({{h}_{k-1}}+{{\tau }_{k-1}}) \\ 
			& \le {\zeta}^{k+1}{{C}_{4}}^{\frac{k}{2}}{{e}^{-{{C}_{1}}\left| \Theta (t) \right|}}{{e}^{{{C}_{2}}\left| \Omega (t) \right|}}V_{1}(0).
		\end{aligned}
	\end{equation}
	According to the DoS attack model, we have $\left| \Theta (t) \right|=t-\left| \Omega (t) \right|$, and further, based on Definition \ref{Def1} and Definition \ref{Def2}, combining \eqref{xuhao32} and \eqref{xuhao33},  it can be obtained
	\begin{eqnarray}\label{xuhao34}
		\begin{aligned}
			V(t)&\le \zeta^{N(0,t)}{{C}_{4}}^{N(0,t)}{{e}^{-{{C}_{1}}(t-\left| \Omega (t) \right|)}}{{e}^{{{C}_{2}}\left| \Omega (t) \right|}}V(0) \\ 
			& ={{e}^{N(0,t)\ln ({\zeta{C}_{4}})}}{{e}^{-{{C}_{1}}t}}{{e}^{({{C}_{1}}+{{C}_{2}})\left| \Omega (t) \right|}}V(0) \\ 
			& ={{e}^{N(0,t)\ln ({\zeta{C}_{4}})}}{{e}^{-{{C}_{1}}t}}{{e}^{({{C}_{1}}+{{C}_{2}})tT}}V(0) \\ 
			& ={{e}^{N(0,t)\ln ({\zeta{C}_{4}})}}{{e}^{[-{{C}_{1}}+({{C}_{1}}+{{C}_{2}})T]t}}V(0).			
		\end{aligned}
	\end{eqnarray}
	The attack frequency is set as $F(0,t)\le \frac{{{k}^{*}}}{\ln ({\zeta{C}_{4}})}$, where ${k}^{*}\in (0,{{C}_{1}})$. Consequently, it can be further derived that
	\begin{eqnarray}\label{xuhao35}
		\begin{aligned}
			V(t)&\le {{e}^{{{k}^{*}}t}}{{e}^{[-{{C}_{1}}+({{C}_{1}}+{{C}_{2}})T]t}}V(0) \\ 
			& ={{e}^{[-{{C}_{1}}+({{C}_{1}}+{{C}_{2}})T+{{k}^{*}}]t}}V(0) \\ 
			& ={{e}^{-{{C}_{5}}t}}V(0).				
		\end{aligned}
	\end{eqnarray}
	Setting the total attack length rate as $T<\frac{{{C}_{1}}-{{k}^{*}}}{{{C}_{1}}+{{C}_{2}}}$, then ${C}_{5}={C}_{1}-({C}_{1}+{C}_{2})T-{k}^{*}>0$.
	
	In conclusion, as $t\to \infty $, $\underset{t\to \infty }{\mathop{\lim }}\,V(t)=0$. Therefore, under DoS attacks, the optimal formation control protocol designed in this paper ensures that the formation error of the nonlinear MAS (1) exponentially converges to zero.
\end{proof}
\begin{remark}  
	Theorem \ref{thm1} proves the stability of the nonlinear MASs \eqref{xuhao1} and \eqref{xuhao2} under the conditions of absence of DoS attack and under DoS attack, When the system is free from DoS attacks, according to the optimal control law based on event triggering, the time derivative of the value function can be obtained as $\nabla V_{i}^{*}<0$,  and because $V_{i}^{*}$ serves as a Lyapunov candidate function for $e_i$, the system is stable. When the system is under attack, the upper bounds for the length rate and frequency of DoS attacks are provided. Eventually, it is proven that the formation error of the system is exponentially stable.	 
\end{remark}
\subsection{Policy Iteration (PI) Algorithm for Coupled HJB Equations}

This subsection will presents the Policy Iteration (PI) algorithm to solve the coupled HJB equations for the agents. The PI algorithm primarily consists of two steps: policy evaluation and policy improvement. The entire process is iteratively executed until the policy converges to the optimal policy, and the corresponding value function also converges to the optimal value function. Additionally, the value function needs to be evaluated based on acceptable control policies. During the execution of the algorithm, initially, initialize the control policies as $u_{1}^{0},\ldots ,u_{N}^{0}$ and the errors as ${e}_{1}(0),\ldots ,{e}_{N}(0)$, subsequently, solve the N-tuple value functions $V_{1}^{\ell },\ldots ,V_{N}^{\ell }$ through the equation ${{H}_{i}}({{e}_{i}},\nabla V_{i}^{\ell },u_{i}^{\ell },u_{j}^{\ell })=0$. Next, update the N-tuple of control policies using $u_{i}^{\ell +1}=-\frac{1}{2}({{d}_{i}}+{{b}_{i}}){{R}_{ii}^{-1}}{{B}^{T}}\nabla V_{i}^{\ell }$, and finally, check for convergence.  The specific algorithmic procedure can be expressed as the algorithm \ref{alg1}.
\begin{algorithm}
	\caption{Policy Iteration for Solving the HJB Equation}\label{alg1}
	\begin{algorithmic}
		\State Initialize Policy:
		\State Step 1: Set initial values ${e}_{1}(0),\ldots,{e}_{N}(0)$ and initial admissible control policies $u_{1}^{0},\ldots,u_{N}^{0}$.
		\State Policy Evaluation:
		\State Step 2: Given the N-tuple of control policies $u_{1}^{\ell },\ldots ,u_{N}^{\ell }$, solve the N-tuple value functions $V_{1}^{\ell },\ldots ,V_{N}^{\ell }$ using the following equations:
		\begin{eqnarray}\label{xuhaog4}
			\begin{aligned}
				{{H}_{i}}({{e}_{i}},\nabla V_{i}^{\ell },u_{i}^{\ell },u_{j}^{\ell })=0.				
			\end{aligned}
		\end{eqnarray}
		\State Policy Improvement:
		\State Step 3: Update the N-tuple of control policies using the following equations,
		\begin{eqnarray}\label{xuhaog5}
			\begin{aligned}
				u_{i}^{\ell +1}=-\frac{1}{2}({{d}_{i}}+{{b}_{i}}){{R}_{ii}}^{-1}{{B}^{T}}\nabla V_{i}^{\ell }.				
			\end{aligned}
		\end{eqnarray}
		\State Check Convergence:
		\State Step 4: If converged, end. Otherwise, set $\ell =\ell +1$ and return to Step 2.    
	\end{algorithmic}
\end{algorithm}

The convergence of the PI algorithm for MASs has been proven in the literature\cite{zhao2019distributed}. However,  due to the nonlinearity of HJB equation, it is difficult to solve. Additionally, the event triggering mechanism \eqref{xuhaogg5} and the PI mechanism are independent, resulting in high computational costs. Therefore, the following Neural Networks approach is employed to solve the HJB equations \eqref{xuhao11}, where the updating of network weights is determined by the event-triggering mechanism \eqref{xuhaogg5}.
\section{Approximate Solutions of Coupled HJB Equations using NNs}\label{sec_4}
This section develops an adaptive algorithm within the framework of Neural Networks for approximating solutions to coupled HJB equations \eqref{xuhao11}.

In order to solve the coupled HJB equation \eqref{xuhao11}, this section employs the NN method to approximate the value functions and compute the optimal control policies. Initially, a critic neural network based on events is constructed to approximate the performance index, the value function ${V}_{i}$ is approximated as
\begin{eqnarray}\label{xuhao36}
	\begin{aligned}
		{{V}_{i}}({{e}_{i}})={{\Gamma }_{ci}^{T}}{{\phi }_{ci}}({{\omega }_{ci}})+{{\varepsilon }_{ci}},				
	\end{aligned}
\end{eqnarray}
where ${{\Gamma }_{ci}}$ represents the target weight matrix from the hidden layer to the output layer, ${{\phi }_{ci}}$ is the activation function, ${{\varepsilon }_{ci}}$ is the NN-based critical approximation error, and ${{\omega }_{ci}}$ is a vector containing ${{e}_{i}}$ and ${{e}_{j}}$. Since the target weight matrix is unknown, the actual output of the critic network can be expressed as:
\begin{eqnarray}\label{xuhaog36}
	\begin{aligned}
		{{\hat{V}}_{i}}({{e}_{i}})={{\hat{\Gamma }}_{ci}}^{T}{{\phi }_{ci}}({{\omega }_{ci}}),				
	\end{aligned}
\end{eqnarray}
where ${{\hat{\Gamma }}_{ci}}$ is an estimate of the target weight matrix ${{\Gamma }_{ci}}$, and ${{\hat{V}}_{i}}({{e}_{i}})$ is an estimate of the target value function.

In the absence of network attacks, the estimated HJB equation is defined as
\begin{equation}\label{xuhao37}
	\begin{aligned}
		&{{H}_{i}}({{e}_{i}},{{{\hat{\Gamma }}}_{ci}},{{u}_{i}},{{u}_{(j)}})\\
		&={{e}_{i}}^{T}{{Q}_{ii}}{{e}_{i}}+\sum\limits_{j\in {{N}_{i}}}{{{u}_{j}}^{T}{{R}_{ij}}{{u}_{j}}}+\left({{{\hat{\Gamma }}}_{ci}}^{T}\nabla {{\phi }_{ci}}({{\omega }_{ci}}) \right) ^{T}\\
		&\;\;\;\; \times\Big(A{{e}_{i}}+{{B}_{1}}{{u}_{i}}-B\sum\limits_{j\in {{N }_{i}}}{{a}_{ij}}{{u}_{j}}+{{F}_{i}}+C\Big)+{{u}_{i}}^{T}{{R}_{ii}}{{u}_{i}} \\ 
		& ={{\delta }_{ci}},			
	\end{aligned}
\end{equation}
where $\nabla {{\phi }_{ci}}({{\omega }_{ci}})= \partial {{\phi }_{ci}}({{\omega }_{ci}})/{\partial {{e}_{i}}}$. Therefore, the critic network can update the parameters ${{\hat{\Gamma }}_{ci}}$ using the gradient descent method to minimize the error function ${{E}_{ci}}=\frac{1}{2}\delta _{ci}^{2}$. Due to the presence of event-triggering mechanisms, neural network weights only need adjustment at event sampling times. The update rule is designed as follows
\begin{eqnarray}\label{xuhao38}
	\left\{
	\begin{aligned}
		{{\dot{\hat{\Gamma }}}_{ci}}&=0,\text{  }t\ne t_{k}^{i},\\
		\hat{\Gamma }_{ci}^{+}&={{\hat{\Gamma }}_{ci}}-{{\alpha }_{ci}}{{\varsigma }_{i}}\Big[{{\varsigma }_{i}}^{T}{{\hat{\Gamma }}_{ci}}+{{e}_{i}}^{T}{{Q}_{ii}}{{e}_{i}}+{{u}_{i}}^{T}{{R}_{ii}}{{u}_{i}} +\sum\limits_{j\in {{N}_{i}}}{{{u}_{j}}^{T}{{R}_{ij}}{{u}_{j}}}\Big],\text{  }t=t_{k}^{i},		
	\end{aligned}
	\right.
\end{eqnarray}
where ${{\varsigma }_{i}}=\nabla {{\phi }_{ci}}\big(A{{e}_{i}}(t)+{{B}_{1}}{{u}_{i}}(t)-B\sum\limits_{j\in {{N }_{i}}}{{a}_{ij}}{{u}_{j}}(t)+{{F}_{i}}(t)+C\big)$, $\alpha _{ci}\in (0,1)$ is the weight learning rate, and $\hat{\Gamma }_{ci}^{+}={{\hat{\Gamma }}_{ci}}(t_{k}^{i})$. Based on the estimate of the value function ${{\hat{V}}_{i}}({{e}_{i}})$ and the estimated weight adjustment  ${{\hat{\Gamma }}_{ci}}$,  the admissible coordinated control policy can be expressed as
\begin{eqnarray}\label{xuhao39}
	\begin{aligned}
		{{\hat{u}}_{i}}=-\frac{1}{2}({{d}_{i}}+{{b}_{i}}){{R}_{ii}}^{-1}{{B}^{T}}\nabla {{\phi }_{ci}^{T}}({{\omega }_{ci}}){{\hat{\Gamma }}_{ci}},\text{  }t=t_{k}^{i}.			
	\end{aligned}
\end{eqnarray}
In order to derive the weight estimation error, substituting \eqref{xuhao36} into \eqref{xuhao8}, we obtain
\begin{equation}\label{xuhao40}
	\begin{aligned}
		&0={{H}_{i}}({{e}_{i}},\nabla {{V}_{i}},{{u}_{i}},{{u}_{(i)}}) \\ 
		& =e_{i}^{T}{{Q}_{ii}}{{e}_{i}}+\sum\limits_{j}{u_{j}^{T}{{R}_{ij}}{{u}_{j}}}+(\frac{\partial {{\phi }_{ci}}({{\omega }_{ci}})^{T}}{\partial {{e}_{i}}}{{{\Gamma }}_{ci}}+\frac{\partial {{\varepsilon }_{ci}}}{\partial {{e}_{i}}})({{{\dot{e}}}_{i}}) \\ 
		& =e_{i}^{T}{{Q}_{ii}}{{e}_{i}}+\sum\limits_{j}{u_{j}^{T}{{R}_{ij}}{{u}_{j}}}+\nabla {{\phi }_{ci}^{T}}({{\omega }_{ci}}){{{\Gamma }}_{ci}}(A{{e}_{i}}+{{B}_{1}}{{u}_{i}}-B\sum\limits_{j\in {{N }_{i}}}{{a}_{ij}}{{u}_{j}}+{{F}_{i}}+C)\\
		&\;\;\;\; +\frac{\partial {{\varepsilon }_{ci}}}{\partial {{e}_{i}}}(A{{e}_{i}}+{{B}_{1}}{{u}_{i}}-B\sum\limits_{j\in {{N }_{i}}}{{a}_{ij}}{{u}_{j}}+{{F}_{i}}+C).			
	\end{aligned}
\end{equation}
Further, rearranging terms, we get
\begin{eqnarray}\label{xuhao41}
	\begin{aligned}
		& e_{i}^{T}{{Q}_{ii}}{{e}_{i}}+\sum\limits_{j}{u_{j}^{T}{{R}_{ij}}{{u}_{j}}}+\nabla {{\phi }_{ci}^{T}}({{\omega }_{ci}}){{{\Gamma }}_{ci}}(A{{e}_{i}}+{{B}_{1}}{{u}_{i}}-B\sum\limits_{j\in {{N }_{i}}}{{a}_{ij}}{{u}_{j}}+{{F}_{i}}+C) \\
		& =-\frac{\partial {{\varepsilon }_{ci}}}{\partial {{e}_{i}}}(A{{e}_{i}}+{{B}_{1}}{{u}_{i}}-B\sum\limits_{j\in {{N }_{i}}}{{a}_{ij}}{{u}_{j}}+{{F}_{i}}+C) \\ 
		& ={{\varepsilon }_{H{{J}_{i}}}},			
	\end{aligned}
\end{eqnarray}
where ${{\varepsilon }_{H{{J}_{i}}}}$ represents the residual caused by function approximation error. Defining the weight estimation error as ${{\tilde{\Gamma }}_{ci}}={{\hat{\Gamma }}_{ci}}-{{\Gamma }_{ci}}$, we can further derive
\begin{enumerate}
	\item [(1)] When the event-triggering condition is not satisfied, i.e., $\text{ }t\in [t_{k}^{i},t_{k+1}^{i})$.
	\begin{eqnarray}\label{xuhao42}
		\begin{aligned}
			& {{{\dot{\tilde{\Gamma }}}}_{ci}}={{{\dot{\hat{\Gamma }}}}_{ci}}-{{{\dot{\Gamma }}}_{ci}} =0.			
		\end{aligned}
	\end{eqnarray}
	\item [(2)] When the event-triggering condition is satisfied, i.e., $\text{ }t=t_{k}^{i}$.
	\begin{equation}\label{xuhao43}
		\begin{aligned}
			\tilde{\Gamma }_{ci}^{+}&={{{\tilde{\Gamma }}}_{ci}}-{{\alpha }_{ci}}{{\varsigma }_{i}}\bigg[{{\varsigma }_{i}}^{T}{{{\hat{\Gamma }}}_{ci}}+{{e}_{i}}^{T}{{Q}_{ii}}{{e}_{i}}+{{u}_{i}}^{T}{{R}_{ii}}{{u}_{i}} +\sum\limits_{j\in {{N}_{i}}}{{{u}_{j}}^{T}{{R}_{ij}}{{u}_{j}}}\bigg] \\ 
			& ={{{\tilde{\Gamma }}}_{ci}}-{{\alpha }_{ci}}{{\varsigma }_{i}}[{{\varsigma }_{i}}^{T}{{{\hat{\Gamma }}}_{ci}}+{{\varepsilon }_{H{{J}_{i}}}}-{{\varsigma }_{i}}^{T}{{\Gamma }_{ci}}] \\ 
			& ={{{\tilde{\Gamma }}}_{ci}}-{{\alpha }_{ci}}{{\varsigma }_{i}}[{{\varsigma }_{i}}^{T}{{{\tilde{\Gamma }}}_{ci}}+{{\varepsilon }_{H{{J}_{i}}}}].			
		\end{aligned}
	\end{equation}
\end{enumerate}
To analyze the uniform boundedness of the weight estimation error, the following assumption is introduced.
\begin{Assumption}\label{Assu3}\cite{zhang2014leader}
	The PE (persistent excitation) condition ensures that $\underline{\varsigma}_{i} \le \Vert {{\varsigma }_{i}} \Vert \le {{\bar{\varsigma }}_{i}}$, where $\underline{\varsigma}_{i}$ and ${{\bar{\varsigma }}_{i}}$ are positive constants.
\end{Assumption}
\begin{Assumption}\label{Assu4}\cite{zhang2014leader}
	The weight vector ${{\Gamma }_{ci}}$, activation function ${{\phi }_{ci}}({{\omega }_{ci}})$, and error vector ${{\varepsilon }_{H{{J}_{i}}}}$ are all bounded, satisfying $\underline{{\Gamma }}_{ci} \le \left\| {{\Gamma }_{ci}} \right\|\le {{\bar{\Gamma }}_{ci}}$, $\left\| {{\phi }_{ci}}({{\omega }_{ci}}) \right\|\le {{\bar{\phi }}_{ci}}$, $\left\| {{\varepsilon }_{H{{J}_{i}}}} \right\|\le {{\bar{\varepsilon }}_{H{{J}_{i}}}}$, where $\underline{{\Gamma }}_{ci}$, ${{\bar{\Gamma }}_{ci}}$, ${{\bar{\phi }}_{ci}}$ and ${{\bar{\varepsilon }}_{H{{J}_{i}}}}$ are all positive constants.
\end{Assumption}
Next, the following theorem is presented to prove that the weight estimation error remains  ultimately uniformly bounded.
\begin{theorem}\label{thm2}
	Consider a  nonlinear MAS \eqref{xuhao1} and \eqref{xuhao2} satisfying Assumptions \ref{Assu1} - \ref{Assu4}, under the weight update law \eqref{xuhao38} and admissible coordinated control policy \eqref{xuhao39}. If the event-triggering condition \eqref{xuhao015} is satisfied, updating the weight parameters ${{\hat{\Gamma }}_{ci}}$ and control policy ${{\hat{u}}_{i}}$, then the weight estimation error ${{\tilde{\Gamma }}_{ci}}$ is ultimately uniformly bounded. 
\end{theorem}
\begin{proof}
	When $\text{ }t\in [t_{k}^{i},t_{k+1}^{i})$,  the following Lyapunov candidate function is constructed
	\begin{eqnarray}\label{xuhao44}
		\begin{aligned}
			{{L}_{i}}={{L}_{1i}}+{{L}_{2i}},			
		\end{aligned}
	\end{eqnarray}
	where ${{L}_{1i}}={{\kappa }_{i}}{{V}_{i}}({{e}_{i}})$, ${{L}_{2i}}=\frac{\operatorname{tr}({{{\tilde{\Gamma }}}_{ci}}^{T}{{{\tilde{\Gamma }}}_{ci}})}{2{{\alpha }_{ci}}}$, ${{\kappa }_{i}}>0$.
	Differentiating  $L_{1i}$ and $L_{2i}$ with respect to time, respectively, we have
	\begin{equation}\label{xuhao45}
		\begin{aligned}
			{{{\dot{L}}}_{1i}}&={{\kappa }_{i}}\nabla {{V}_{i}}\\
			&=-{{\kappa }_{i}}\sum\limits_{i=1}^{N}{\bigg[ e_{i}^{T}{{Q}_{ii}}{{e}_{i}}+\sum\limits_{j}{u_{j}^{T}{{R}_{ij}}{{u}_{j}}} \bigg]} \\ 
			& \le -\sum\limits_{i=1}^{N}{{{\kappa }_{i}}{{\lambda }_{\min }}({{Q}_{ii}}){{\Vert {{e}_{i}} \Vert}^{2}}}-\sum\limits_{i=1}^{N}{{{\sum\limits_{j}{{{\kappa }_{i}}{{\lambda }_{\min }}({{R}_{ij}})\Vert {{u}_{j}} \Vert}}^{2}}},			
		\end{aligned}
	\end{equation}
	and
	\begin{eqnarray}\label{xuhao46}
		\begin{aligned}
			{{\dot{L}}_{2i}}=0,		
		\end{aligned}
	\end{eqnarray}
	Therefore, we obtain
	\begin{equation}
		\begin{aligned}
			{{\dot{L}}_{i}}&={{\dot{L}}_{1i}}+{{\dot{L}}_{2i}}\\
			&\le -\sum\limits_{i=1}^{N}{{{\kappa }_{i}}{{\lambda }_{\min }}({{Q}_{ii}}){{\Vert {{e}_{i}} \Vert}^{2}}}-\sum\limits_{i=1}^{N}{{{\sum\limits_{j}{{{\kappa }_{i}}{{\lambda }_{\min }}({{R}_{ij}})\Vert {{u}_{j}} \Vert}}^{2}}}\\
			&\le 0.
		\end{aligned}
	\end{equation}
	According to Lyapunov stability theory, the weight estimation error ${{\tilde{\Gamma }}_{ci}}$  is ultimately uniformly bounded.\\
	When $\text{ }t=t_{k}^{i}$, construct the Lyapunov candidate function as
	\begin{eqnarray}\label{xuhao47}
		\begin{aligned}
			{{L}_{i}}={{L}_{3i}}+{{L}_{4i}},			
		\end{aligned}
	\end{eqnarray}
	where ${{L}_{3i}}={{\kappa }_{i}}{{V}_{i}}({{e}_{i}}^{+})-{{\kappa }_{i}}{{V}_{i}}({{e}_{i}})$, ${{L}_{4i}}=\frac{\operatorname{tr}({{{\tilde{\Gamma }}}_{ci}}^{+T}{{{\tilde{\Gamma }}}_{ci}}^{+})}{2{{\alpha }_{ci}}}-\frac{\operatorname{tr}({{{\tilde{\Gamma }}}_{ci}}^{T}{{{\tilde{\Gamma }}}_{ci}})}{2{{\alpha }_{ci}}}$. Since ${{\dot{L}}_{1i}}\le 0$, it can be derived that
	\begin{eqnarray}\label{xuhao48}
		\begin{aligned}
			{{L}_{3i}}={{\kappa }_{i}}{{V}_{i}}({{e}_{i}}^{+})-{{\kappa }_{i}}{{V}_{i}}({{e}_{i}})\le 0.			
		\end{aligned}
	\end{eqnarray}
	According to equation \eqref{xuhao43}, we have
	\begin{equation}\label{xuhao49}
		\begin{aligned}
			{{L}_{4i}}&=\frac{\operatorname{tr}({{{\tilde{\Gamma }}}_{ci}}^{+T}{{{\tilde{\Gamma }}}_{ci}}^{+})}{2{{\alpha }_{ci}}}-\frac{\operatorname{tr}({{{\tilde{\Gamma }}}_{ci}}^{T}{{{\tilde{\Gamma }}}_{ci}})}{2{{\alpha }_{ci}}} \\ 
			& =\frac{\operatorname{tr}{{\left[ {{{\tilde{\Gamma }}}_{ci}}-{{\alpha }_{ci}}{{\varsigma }_{i}}[{{\varsigma }_{i}}^{T}{{{\tilde{\Gamma }}}_{ci}}+{{\varepsilon }_{H{{J}_{i}}}}] \right]}^{T}}\left[ {{{\tilde{\Gamma }}}_{ci}}-{{\alpha }_{ci}}{{\varsigma }_{i}}[{{\varsigma }_{i}}^{T}{{{\tilde{\Gamma }}}_{ci}}+{{\varepsilon }_{H{{J}_{i}}}}] \right]}{2{{\alpha }_{ci}}}-\frac{\operatorname{tr}({{{\tilde{\Gamma }}}_{ci}}^{T}{{{\tilde{\Gamma }}}_{ci}})}{2{{\alpha }_{ci}}} \\ 
			& =-\operatorname{tr}\left[ {{{\tilde{\Gamma }}}_{ci}}^{T}{{\varsigma }_{i}}[{{\varsigma }_{i}}^{T}{{{\tilde{\Gamma }}}_{ci}}+{{\varepsilon }_{H{{J}_{i}}}}] \right]+\frac{{{\alpha }_{ci}}}{2}\operatorname{tr}{{[{{\varsigma }_{i}}^{T}{{{\tilde{\Gamma }}}_{ci}}+{{\varepsilon }_{H{{J}_{i}}}}]}^{T}}{{\varsigma }_{i}}^{T}{{\varsigma }_{i}}[{{\varsigma }_{i}}^{T}{{{\tilde{\Gamma }}}_{ci}}+{{\varepsilon }_{H{{J}_{i}}}}] \\ 
			& =-\operatorname{tr}\left[ {{{\tilde{\Gamma }}}_{ci}}^{T}{{\varsigma }_{i}}{{\varsigma }_{i}}^{T}{{{\tilde{\Gamma }}}_{ci}} \right]-\operatorname{tr}\left[ {{{\tilde{\Gamma }}}_{ci}}^{T}{{\varsigma }_{i}}{{\varepsilon }_{H{{J}_{i}}}} \right]+\frac{{{\alpha }_{ci}}}{2}\operatorname{tr}[\tilde{\Gamma }_{ci}^{T}{{\varsigma }_{i}}{{\varsigma }_{i}}^{T}{{\varsigma }_{i}}{{\varsigma }_{i}}^{T}{{{\tilde{\Gamma }}}_{ci}}]\\
			&\;\;\;\;+{{\alpha }_{ci}}\operatorname{tr}[\tilde{\Gamma }_{ci}^{T}{{\varsigma }_{i}}{{\varsigma }_{i}}^{T}{{\varsigma }_{i}}{{\varepsilon }_{H{{J}_{i}}}}] \text{ }+\frac{{{\alpha }_{ci}}}{2}\operatorname{tr}[{{\varepsilon }_{H{{J}_{i}}}}^{T}{{\varsigma }_{i}}^{T}{{\varsigma }_{i}}{{\varepsilon }_{H{{J}_{i}}}}].			
		\end{aligned}
	\end{equation}
	Due to ${{\tilde{\Gamma }}_{ci}}{{\varsigma }_{i}}{{\varsigma }_{i}}^{T}{{\tilde{\Gamma }}_{ci}}>0$ and ${{\varepsilon }_{H{{J}_{i}}}}^{T}{{\varsigma }_{i}}^{T}{{\varsigma }_{i}}{{\varepsilon }_{H{{J}_{i}}}}>0$, there exist ${{m}_{i1}}>0$, ${{m}_{i2}}>0$ such that 
	\begin{equation}
		\begin{aligned}
			{{m}_{i1}}{{\left\| {{{\tilde{\Gamma }}}_{ci}} \right\|}^{2}}\le {{\tilde{\Gamma }}_{ci}}{{\varsigma }_{i}}{{\varsigma }_{i}}^{T}{{\tilde{\Gamma }}_{ci}}, \;\; {{\varepsilon }_{H{{J}_{i}}}}^{T}{{\varsigma }_{i}}^{T}{{\varsigma }_{i}}{{\varepsilon }_{H{{J}_{i}}}}\le {{m}_{i2}}{{\bar{\varepsilon }}_{H{{J}_{i}}}}^{2}.		
		\end{aligned}
	\end{equation}
	Therefore, equation \eqref{xuhao49} can be further estimated
	\begin{equation}\label{xuhao50}
		\begin{aligned}
			{{L}_{4i}}&\le -{{m}_{i1}}{{\left\| {{{\tilde{\Gamma }}}_{ci}} \right\|}^{2}}+\frac{{{\alpha }_{ci}}}{2}{{\left\| {{\varsigma }_{i}} \right\|}^{2}}{{\left\| {{{\tilde{\Gamma }}}_{ci}} \right\|}^{2}}+\frac{2}{{{\alpha }_{ci}}}{{{\bar{\varepsilon }}}_{H{{J}_{i}}}}^{2}+\frac{{{\alpha }_{ci}}}{2}{{m}_{i1}}^{2}{{\left\| {{{\tilde{\Gamma }}}_{ci}} \right\|}^{2}}\\
			&\;\;\;\; +{{\alpha }_{ci}}{{m}_{i1}}{{\left\| {{\varsigma }_{i}} \right\|}^{2}}{{\left\| {{{\tilde{\Gamma }}}_{ci}} \right\|}^{2}}+{{\alpha }_{ci}}{{m}_{i2}}{{{\bar{\varepsilon }}}_{H{{J}_{i}}}}^{2}+\frac{{{\alpha }_{ci}}}{2}{{m}_{i2}}{{{\bar{\varepsilon }}}_{H{{J}_{i}}}}^{2} \\ 
			& \le \left( -{{m}_{i1}}+\frac{{{\alpha }_{ci}}}{2}{{{\bar{\varsigma }}}_{i}}^{2}+\frac{{{\alpha }_{ci}}}{2}{{m}_{i1}}^{2}+{{\alpha }_{ci}}{{m}_{i1}}{{{\bar{\varsigma }}}_{i}}^{2} \right){{\left\| {{{\tilde{\Gamma }}}_{ci}} \right\|}^{2}} +\left( \frac{2}{{{\alpha }_{ci}}}+\frac{3}{2}{{\alpha }_{ci}}{{m}_{i2}} \right){{{\bar{\varepsilon }}}_{H{{J}_{i}}}}^{2}.		
		\end{aligned}
	\end{equation}
	Setting ${{\alpha }_{ci}}\le \frac{2{{m}_{i1}}}{{{{\bar{\varsigma }}}_{i}}^{2}+{{m}_{i1}}^{2}+2{{m}_{i1}}{{{\bar{\varsigma }}}_{i}}^{2}}$, and for sufficiently large $\left\| {{{\tilde{\Gamma }}}_{ci}} \right\|$, the following inequalities hold
	\begin{equation}\label{xuhao51}
		\begin{aligned}
			\left\| {{{\tilde{\Gamma }}}_{ci}} \right\|\ge \frac{4+3{{\alpha }_{ci}}^{2}{{m}_{i2}}}{2{{\alpha }_{ci}}{{m}_{i1}}-{{\alpha }_{ci}}^{2}{{{\bar{\varsigma }}}_{i}}^{2}-{{\alpha }_{ci}}^{2}{{m}_{i1}}^{2}-2{{\alpha }_{ci}}^{2}{{m}_{i1}}{{{\bar{\varsigma }}}_{i}}^{2}}{{\bar{\varepsilon }}_{H{{J}_{i}}}}.	
		\end{aligned}
	\end{equation}
	Therefore, it can be concluded that ${{L}_{4i}}\le 0$, and further, ${{L}_{i}}={{L}_{3i}}+{{L}_{4i}}\le 0$. According to Lyapunov stability theory, the weight estimation error ${\tilde{\Gamma}_{ci}}$ is uniformly ultimately bounded.
	
	Next, we will demonstrate the convergence of the error between the approximate coordinated control input ${{\hat{u}}_{i}}$ and the optimal coordinated control input $u_{i}^{*}$ to a small neighborhood. Based on (\ref{xuhao36}) and (\ref{xuhao10}), we have
	\begin{eqnarray}\label{xuhao52}
		\begin{aligned}
			{{{\hat{u}}}_{i}}-{{u}_{i}}^{*}&=-\frac{1}{2}({{d}_{i}}+{{b}_{i}}){{R}_{ii}}^{-1}{{B}^{T}}(\nabla {{{\hat{V}}}_{i}}-\nabla V_{i}^{*}) \\ 
			& =-\frac{1}{2}({{d}_{i}}+{{b}_{i}}){{R}_{ii}}^{-1}{{B}^{T}}(\nabla {{\phi }_{ci}^{T}}({{\omega }_{ci}}){{{\hat{\Gamma }}}_{ci}}-\nabla{{\phi }_{ci}^{T}}({{\omega }_{ci}}){{\Gamma }_{ci}}-\nabla {{\varepsilon }_{ci}}) \\ 
			& =-\frac{1}{2}({{d}_{i}}+{{b}_{i}}){{R}_{ii}}^{-1}{{B}^{T}}(\nabla {{\phi }_{ci}^{T}}({{\omega }_{ci}}){{{\tilde{\Gamma }}}_{ci}}-\nabla {{\varepsilon }_{ci}}),
		\end{aligned}
	\end{eqnarray}
	where $\nabla {{\varepsilon }_{ci}} = {\partial{\varepsilon }_{ci}}/{\partial {{e}_{i}}}$. When $t\to \infty $, further conclusions can be drawn
	\begin{eqnarray}\label{xuhao53}
		\begin{aligned}
			\left\| {{{\hat{u}}}_{i}}-{{u}_{i}}^{*} \right\|&=\frac{1}{2}\left\| ({{d}_{i}}+{{b}_{i}}){{R}_{ii}}^{-1}{{B}^{T}}\left(\nabla {{\phi }_{ci}^{T}}({{\omega }_{ci}}){{{\tilde{\Gamma }}}_{ci}}-\nabla {{\varepsilon }_{ci}}\right) \right\| \\ 
			& \le \frac{1}{2}\left\| ({{d}_{i}}+{{b}_{i}}){{R}_{ii}}^{-1}{{B}^{T}} \right\|\left\| \left(\nabla {{\phi }_{ci}^{T}}({{\omega }_{ci}}){{{\tilde{\Gamma }}}_{ci}}-\nabla {{\varepsilon }_{ci}}\right) \right\| \\ 
			& \le \frac{1}{2}\left\| ({{d}_{i}}+{{b}_{i}}){{R}_{ii}}^{-1}{{B}^{T}} \right\|\left( \left\| {{{\tilde{\Gamma }}}_{ci}} \right\|\left\| \nabla {{\phi }_{ci}}({{\omega }_{ci}}) \right\|+\left\| \nabla {{\varepsilon }_{ci}}) \right\| \right).
		\end{aligned}
	\end{eqnarray}
	Due to the boundedness of ${{\varsigma }_{i}}$ and ${{\varepsilon }_{H{{J}_{i}}}}$, both $\left\| \nabla {{\phi }_{ci}}({{\omega }_{ci}}) \right\|$ and $\left\| \nabla {{\varepsilon }_{ci}}) \right\|$ are bounded. Furthermore, based on the previous proof, it is established that $\left\| {{{\tilde{\Gamma }}}_{ci}} \right\|$ is also bounded. Therefore, the error between the approximate coordinated control input ${{\hat{u}}_{i}}$ and the optimal coordinated control input $u_{i}^{*}$ converges to a small neighborhood. 	
\end{proof}
\begin{figure}[htbp]
\centering{\includegraphics[width=12cm]{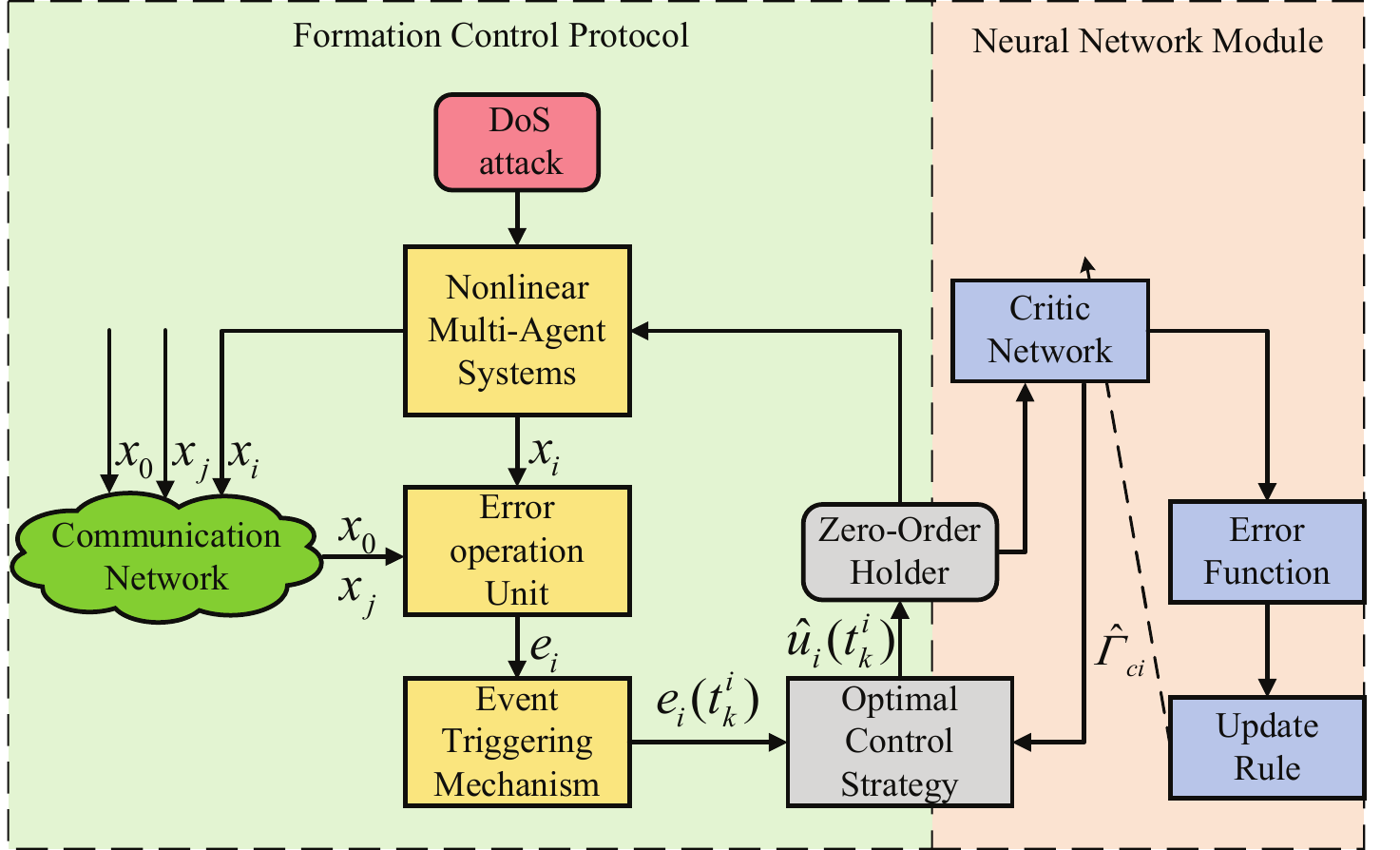}}
\caption{Formation tracking controller for the $i$th follower }
\label{liuchengtu}
\end{figure}
The proposed event-triggered optimal formation control method based on ADP is illustrated in Figure \ref{liuchengtu}. By integrating ADP with event-triggered control methods, the weight parameters of the neural network are updated only at event-triggering instants, effectively reducing computational burden.

Next, the following theorem is provided to demonstrate that the proposed event-triggering mechanism \eqref{xuhaogg5} in this paper does not exhibit Zeno behavior.
\begin{theorem}\label{thm3}
	For the nonlinear MAS \eqref{xuhao1} and \eqref{xuhao2} satisfying Assumptions \ref{Assu1} - \ref{Assu4}, given the event-triggering condition \eqref{xuhao015}, the time interval between events $t_{k+1}^{i}-t_{k}^{i}$ has a positive lower bound.
\end{theorem}
\begin{proof}
	According to equation \eqref{xuhao4}, we have
	\begin{equation}\label{xuhao54}
		\begin{aligned}
			&\left\| {{{\dot{\delta }}}_{i}}(t) \right\|= \left\| {{{\dot{e}}}_{i}}(t) \right\| \\ 
			& =\left\| A{{e}_{i}}(t)+{{B}_{1}}{{u}_{i}}(t)-B\sum\limits_{j\in {{N }_{i}}}{{a}_{ij}}{{u}_{j}}(t)+F_{i}(t)+C \right\| \\ 
			& \le \left\| A{{e}_{i}}(t) \right\|+\left\| {{B}_{1}}{{u}_{i}}(t) \right\|+\left\|B\sum\limits_{j\in {{N }_{i}}}{{a}_{ij}}{{u}_{j}}(t)\right\| +\lambda \left\| {{e}_{i}}(t) \right\|+C \\ 
			& \le {{C}_{6}}\left\| {{e}_{i}}(t_{k}^{i})-{{\delta }_{i}}(t) \right\|+\left\| {{B}_{1}}{{u}_{i}}(t)\right\| +\left\|B\sum\limits_{j\in {{N }_{i}}}{{a}_{ij}}{{u}_{j}}(t)\right\| +C \\ 
			& \le {{C}_{6}}\left\| {{\delta }_{i}}(t) \right\|+{{C}_{6}}\left\| {{e}_{i}}(t_{k}^{i}) \right\| +\left\| \frac{1}{2}{{B}_{1}}({{d}_{i}}+{{b}_{i}}){{R}_{ii}}^{-1}{{B}^{T}}\nabla {{\phi }_{ci}^{T}}({{\omega }_{ci}}){{\hat{\Gamma }}_{ci}} \right\|\\
			&\;\;\;\; +\left\|\frac{1}{2}B\sum\limits_{j\in {{N }_{i}}}{{a}_{ij}}({{d}_{j}}+{{b}_{j}}){{R}_{ij}}^{-1}{{B}^{T}}\nabla {{\phi }_{cj}^{T}}({{\omega }_{cj}}){{\hat{\Gamma }}_{cj}}\right\|+C \\ 
			& \le {{C}_{6}}\left\| {{\delta }_{i}}(t) \right\|+{{C}_{7}},
		\end{aligned}
	\end{equation}
	where ${{C}_{6}}={{\lambda }_{\max }}(A)+\lambda $,\\ ${{C}_{7}}=\left\| \frac{1}{2}{{B}_{1}}({{d}_{i}}+{{b}_{i}}){{R}_{ii}}^{-1}{{B}^{T}}\nabla {{\phi }_{ci}^{T}}({{\omega }_{ci}}){{\hat{\Gamma }}_{ci}} \right\|
	+\left\|\frac{1}{2}B\sum\limits_{j\in {{N }_{i}}}{{a}_{ij}}({{d}_{j}}+{{b}_{j}}){{R}_{ij}}^{-1}{{B}^{T}}\nabla {{\phi }_{cj}^{T}}({{\omega }_{cj}}){{\hat{\Gamma }}_{cj}}\right\|+{{C}_{6}}\left\| {{e}_{i}}(t_{k}^{i}) \right\|+C$.
	Based on Theorem \ref{thm2}, since ${{\hat{\Gamma }}_{ci}}$ and ${{\hat{\Gamma }}_{cj}}$ are bounded, ${{C}_{7}}$ is also bounded. Furthermore, it can be derived
	\begin{eqnarray}\label{xuhao55}
		\begin{aligned}
			\left\| {{\delta }_{i}}(t) \right\|\le {{e}^{{{C}_{6}}(t-t_{k}^{i+})}}{{\delta }_{i}}(t_{k}^{i+})+\frac{1}{2}\int_{t_{k}^{i+}}^{t}{{{e}^{{{C}_{6}}(t-s)}}C{}_{7}ds}.
		\end{aligned}
	\end{eqnarray}
	When $t=t_{k}^{i}$, the event-triggering error is zero, i.e., ${{\delta }_{i}}(t_{k}^{i+})=0$, then we can obtain
	\begin{eqnarray}\label{xuhao56}
		\begin{aligned}
			\left\| {{\delta }_{i}}(t) \right\|\le \frac{1}{2{{C}_{6}}}{{C}_{7}}({{e}^{{{C}_{6}}(t-t_{k}^{i+})}}-1),\text{   }t\in (t_{k}^{i},t_{k+1}^{i}].
		\end{aligned}
	\end{eqnarray}
	According to the event-triggering function, we can obtain
	\begin{equation}\label{xuhao57}
		\begin{aligned}
			&{{\left\| {{\delta }_{i}} \right\|}^{2}}\le \frac{\sum\limits_{j\in {{N}_{i}}}{{{\lambda }_{\min }}({{R}_{ij}}){{\Vert u_{j}^{*} \Vert}^{2}}+{{\lambda }_{\min }}({{R}_{ii}}){{\Vert u_{i}^{*}(t_{k}^{i}) \Vert}^{2}}}}{\lambda _{\max }^{2}({{R}_{ii}}){{M}^{2}}} \\ 
			& \le \frac{\sum\limits_{j\in {{N}_{i}}}{{{\lambda }_{\min }}({{R}_{ij}}){{\left\| \frac{1}{2}({{d}_{j}}+{{b}_{j}}){{R}_{ij}}^{-1}{{B}^{T}}\nabla {{\phi }_{cj}^{T}}({{\omega }_{ci}}){{\hat{\Gamma }}_{cj}} \right\|}^{2}}}}{\lambda _{\max }^{2}({{R}_{ii}}){{M}^{2}}} \\
			&\;\;\;\; +\frac{{{\lambda }_{\min }}({{R}_{ii}}){{\left\| \frac{1}{2}({{d}_{i}}+{{b}_{i}}){{R}_{ii}}^{-1}{{B}^{T}}\nabla {{\phi }_{ci}^{T}}({{\omega }_{ci}}){{\hat{\Gamma }}_{ci}} \right\|}^{2}}}{\lambda _{\max }^{2}({{R}_{ii}}){{M}^{2}}} \\
			& \le \frac{\sum\limits_{j\in {{N}_{i}}}{{{\lambda }_{\min }}({{R}_{ij}})\Psi_{cj}}}{4\lambda _{\min }^{2}({{R}_{ij}})\lambda _{\max }^{2}({{R}_{ii}}){{M}^{2}}}+\frac{{{\lambda }_{\min }}({{R}_{ii}}){\Psi_{ci}}}{4\lambda _{\min }^{2}({{R}_{ii}})\lambda _{\max }^{2}({{R}_{ii}}){{M}^{2}}}.
		\end{aligned}
	\end{equation}
	where, $\Psi_{cj}={{\left\|({{d}_{j}}+{{b}_{j}}){{B}^{T}}\nabla {{\phi }_{cj}^{T}}({{\omega }_{ci}}){{\hat{\Gamma }}_{cj}} \right\|}^{2}}$. So, at $t=t_{k+1}^{i}$,
	\begin{equation}\label{xuhao58}
		\begin{aligned}
			&\left\| {{\delta }_{i}}(t_{k+1}^{i}) \right\|\le \sqrt{\frac{\sum\limits_{j\in {{N}_{i}}}{{{\lambda }_{\min }}({{R}_{ij}})\Psi_{cj}}}{4\lambda _{\min }^{2}({{R}_{ij}})\lambda _{\max }^{2}({{R}_{ii}}){{M}^{2}}}+\frac{{{\lambda }_{\min }}({{R}_{ii}}){\Psi_{ci}}}{4\lambda _{\min }^{2}({{R}_{ii}})\lambda _{\max }^{2}({{R}_{ii}}){{M}^{2}}}}.
		\end{aligned}
	\end{equation}
	According to  \eqref{xuhao56} and \eqref{xuhao58}, it can be determined that
	\begin{equation}\label{xuhao59}
		\begin{aligned}
			&\sqrt{\frac{\sum\limits_{j\in {{N}_{i}}}{{{\lambda }_{\min }}({{R}_{ij}})\Psi_{cj}}}{4\lambda _{\min }^{2}({{R}_{ij}})\lambda _{\max }^{2}({{R}_{ii}}){{M}^{2}}}+\frac{{{\lambda }_{\min }}({{R}_{ii}}){\Psi_{ci}}}{4\lambda _{\min }^{2}({{R}_{ii}})\lambda _{\max }^{2}({{R}_{ii}}){{M}^{2}}}}\le \frac{1}{2{{C}_{6}}}{{C}_{7}}({{e}^{{{C}_{6}}(t_{k+1}^{i}-t_{k}^{i})}}-1).
		\end{aligned}
	\end{equation}
	Therefore, we can obtain the time interval for event triggering as
	\begin{equation}\label{xuhao60}
		\begin{aligned}
			&t_{k+1}^{i}-t_{k}^{i}\ge \frac{1}{{{C}_{6}}}\log \frac{{2{C}_{6}}\sqrt{\frac{\sum\limits_{j\in {{N}_{i}}}{{{\lambda }_{\min }}({{R}_{ij}})\Psi_{cj}}}{4\lambda _{\min }^{2}({{R}_{ij}})\lambda _{\max }^{2}({{R}_{ii}}){{M}^{2}}}+\frac{{{\lambda }_{\min }}({{R}_{ii}}){\Psi_{ci}}}{4\lambda _{\min }^{2}({{R}_{ii}})\lambda _{\max }^{2}({{R}_{ii}}){{M}^{2}}}}}{{{C}_{7}}}.
		\end{aligned}
	\end{equation}
	Therefore, the time interval between events $t_{k+1}^{i}-t_{k}^{i}$ has a positive lower bound. That is, Zeno behavior is excluded.
\end{proof}
\begin{remark}  
	Theorem \ref{thm3} proves the feasibility of the designed event-triggering mechanism by demonstrating the existence of a positive lower bound on the time interval between events. Compared to traditional continuous-time control methods, event-triggering control only requires signal updates when trigger conditions are satisfied, thus saving resources and reducing power consumption.
\end{remark}
\begin{figure}[htbp]
	\centering
	\includegraphics[width=0.5\linewidth]{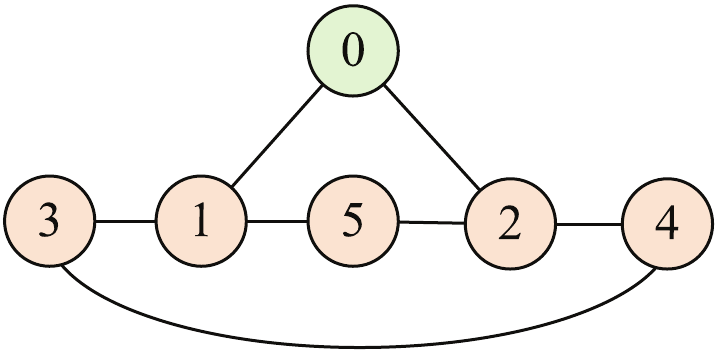}\\
	\caption{The communication topology of the MASs. }\label{fig2}
\end{figure}
\section{Simulation}\label{sec_5}
To validate the effectiveness of the proposed algorithm, numerical simulations are conducted in this section on a MAS consisting of one leader agent and five follower agents. The communication topology of all agents is illustrated in Figure \ref{fig2}. Node 0 represents the leader agent, while the remaining nodes represent the five follower agents. Additionally, there exists a directed path between the leader agent and any follower agent. 
\subsection{Simulation for MAS with formation control}
In this subsection, we will set the parameters for the MAS and analyze the simulation results. For the structure in Figure 2, it can be inferred that the Laplacian matrix is
\begin{equation}\label{xuhao61}
	L=\begin{bmatrix}
		2& 0 &-1  &0  &-1 \\ 
		0& 2 &0  &-1  &-1 \\ 
		-1& 0 &  2&  -1& 0\\ 
		0 & -1 & -1 & 2 &0 \\ 
		-1& -1 & 0 & 0 & 2
	\end{bmatrix}.
\end{equation}
The connection matrix between the leader and followers is
\begin{equation}\label{xuhao62}
	C=\rm{diag}\left \{1,1,0,0,0 \right \}.
\end{equation}
The dynamical model for the $i$th follower is given as follows
\begin{equation}\label{xuhao63}
	{{\dot{x}}_{i}}(t)=0.4\sin (0.1{{x}_{i}}(t))+A{{x}_{i}}(t)+B{{u}_{i}}(t),
\end{equation}
where $A={{I}_{2}}$, $B=0.9{{I}_{2}}$, ${{x}_{i}}=[{x}_{i1},{x}_{i2}]^{T}$.\\
The dynamical model for the leader is represented as
\begin{equation}\label{xuhao64}
	{{\dot{x}}_{0}}(t)=f({{x}_{0}}(t)),
\end{equation}
where $f_{0}({{x}_{0}})=[ 0.7, 0.35\cos ({{x}_{0,1}})+0.2\sin (0.1{{x}_{0,1}}) ]^T$.\\
The initial states for all agents are selected as follows: ${{{x}}_{0}}(0)=[0,0]^T$, ${{{x}}_{1}}(0)=[0.6,3]^T$, ${{{x}}_{2}}(0)=[-0.2,4]^T$, ${{{x}}_{3}}(0)=[-1.5,5]^T$, ${{{x}}_{4}}(0)=[-1.5,0.8]^T$, ${{{x}}_{5}}(0)=[-0.2,1.5]^T$, $i=1,\cdots ,5$. Assume that non-periodic DoS attacks occur at time intervals of $[0.1, 2]$, $[4, 6]$, and $[8, 9]$, choose $\lambda_{\max}(p) = 1.2$, $\lambda_{\min}(p) = 0.8$, $\zeta = 3$, $k^{*} = 0.08$, $C_{1} = 0.4$, $C_{2} = 3.6$, $C_{3}=4$, $C_{4}=4.08$, $F\le \frac{{{k}^{*}}}{\ln ({\zeta{C}_{4}})}=0.0319$,  $T<\frac{{{C}_{1}}-{{k}^{*}}}{{{C}_{1}}+{{C}_{2}}}=0.08$. The cost function matrix is given as ${{Q}_{ii}}=0.4{{I}_{2}}$, ${{R}_{ii}}=0.1$, ${{R}_{ij}}=0.01$. The critic network is a three-layer BP neural network, consisting of 2 input layer neurons, 5 hidden layer neurons, and 1 output layer neuron. Its initial weights are ${{\hat{\Gamma }}_{ci}}={{[0.2,0.46,0.1,0.32,0.61]}^{T}}$, and the learning parameter is chosen as ${{\alpha }_{ci}}=0.1$.
\begin{figure}[htbp]
	\centering
	\includegraphics[width=0.9\linewidth]{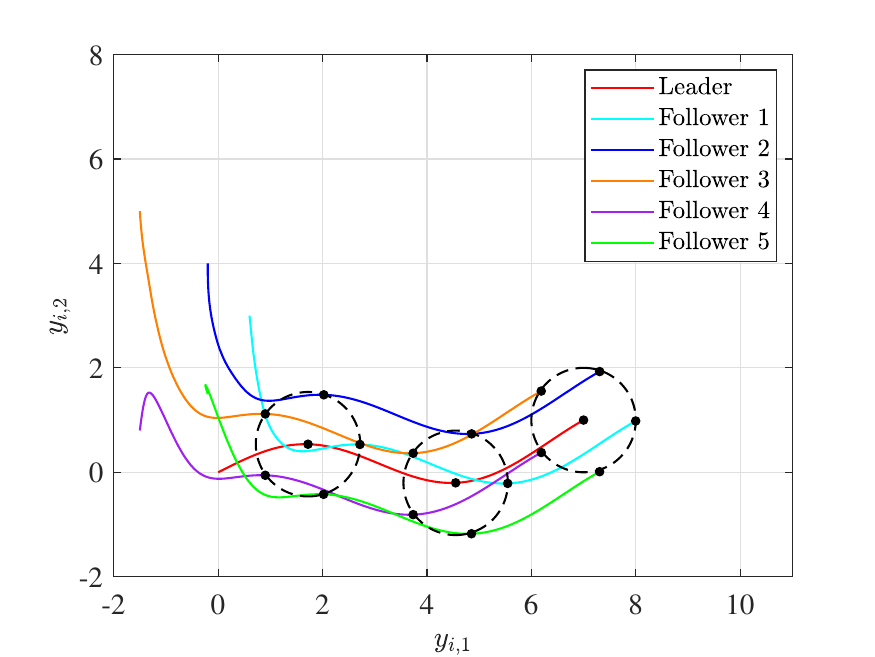}
	\caption{The trajectory of the Multi-agent systems.}
	\label{fig.3}
\end{figure}
\begin{figure}[htbp]
	\centering
	\includegraphics[width=0.9\linewidth]{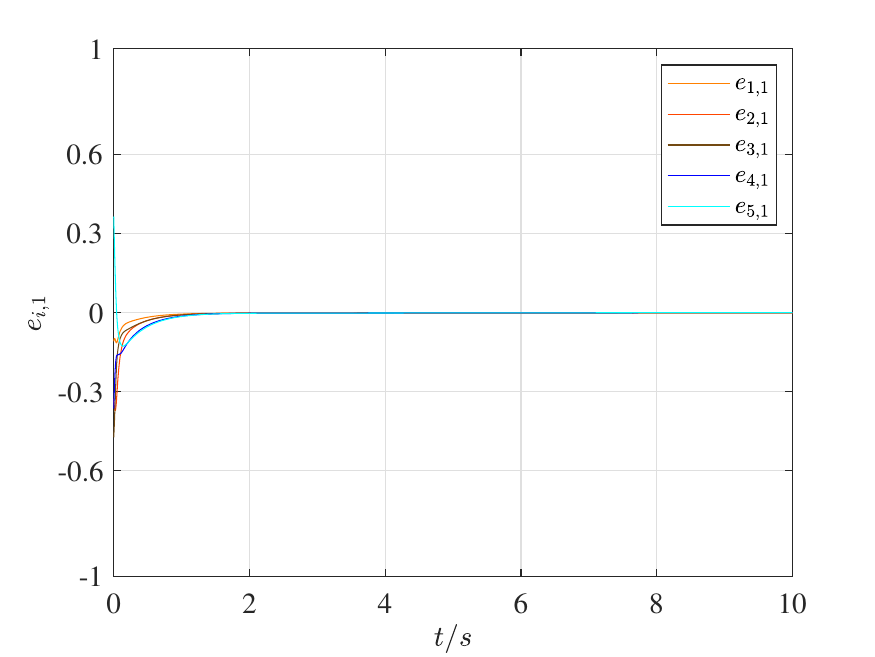}
	\caption{Formation tracking error in horizontal direction.}
	\label{fig.4}
\end{figure}
\begin{figure}[htbp]
	\centering
	\includegraphics[width=0.9\linewidth]{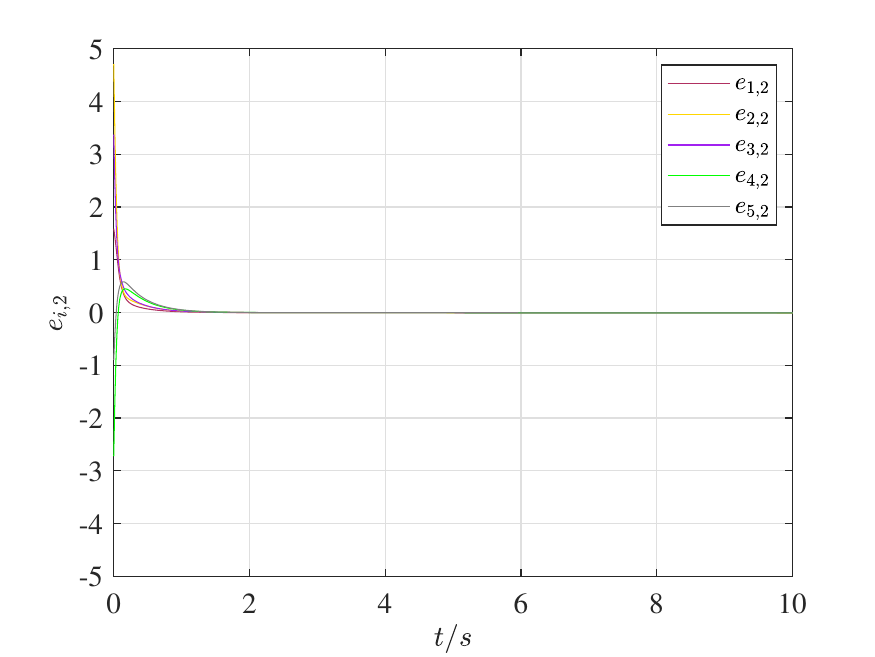}
	\caption{Formation tracking error in vertical direction.}
	\label{fig.5}
\end{figure}
\begin{figure}[htbp]
	\centering
	\includegraphics[width=0.9\linewidth]{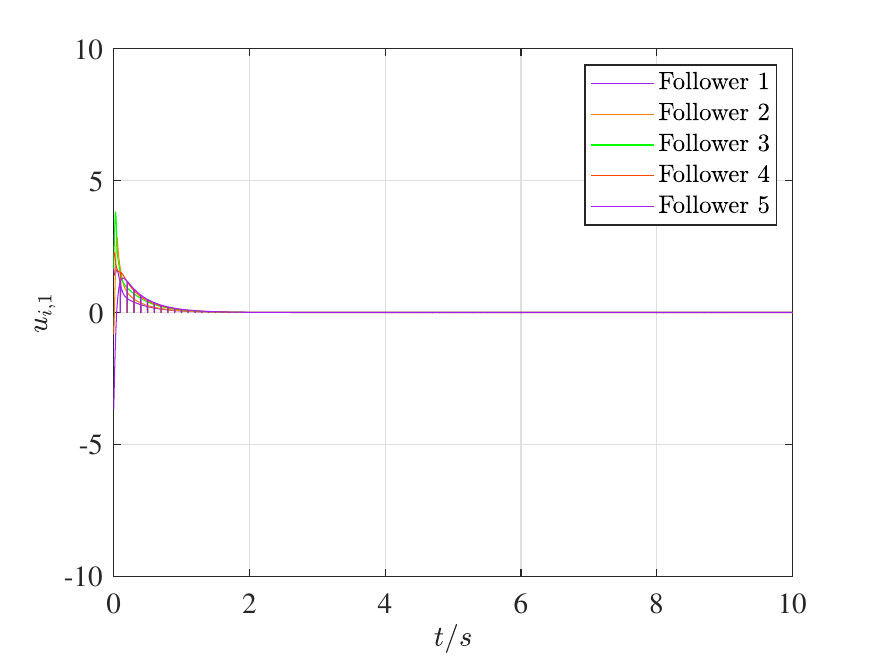}
	\caption{Control input in horizontal direction.}
	\label{fig.6}
\end{figure}
\begin{figure}[htbp]
	\centering
	\includegraphics[width=0.9\linewidth]{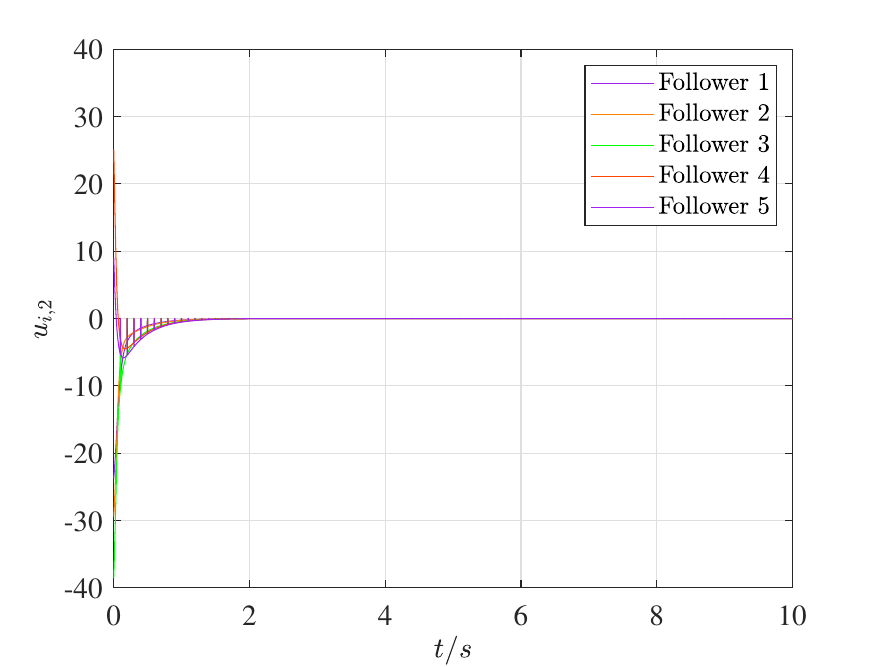}
	\caption{Control input in vertical direction.}
	\label{fig.7}
\end{figure}
\begin{figure}[htbp]
	\centering
	\includegraphics[width=0.9\linewidth]{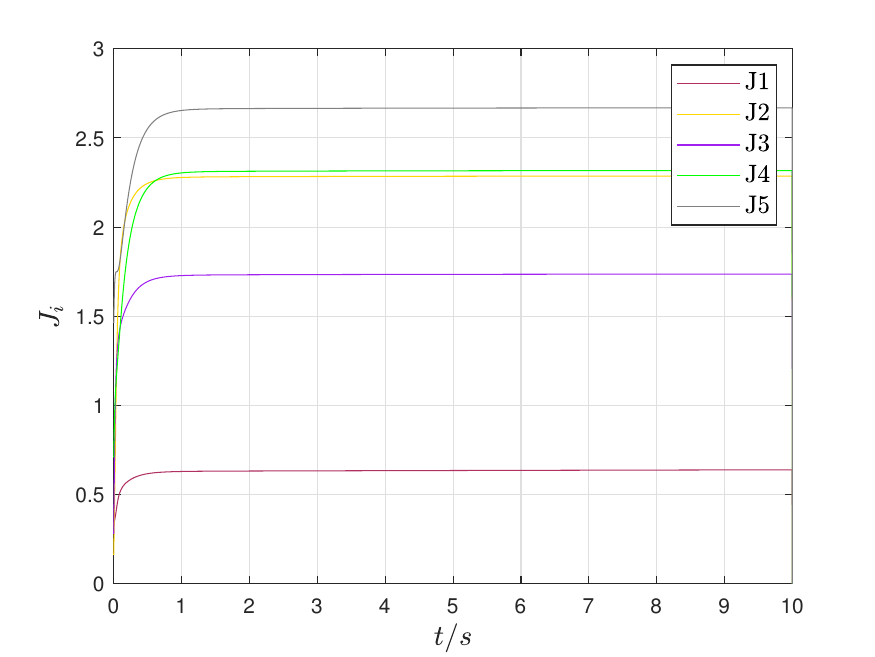}
	\caption{Cumulative costs of all followers.}
	\label{fig.8}
\end{figure}
\begin{figure}[htbp]
	\centering
	\includegraphics[width=0.9\linewidth]{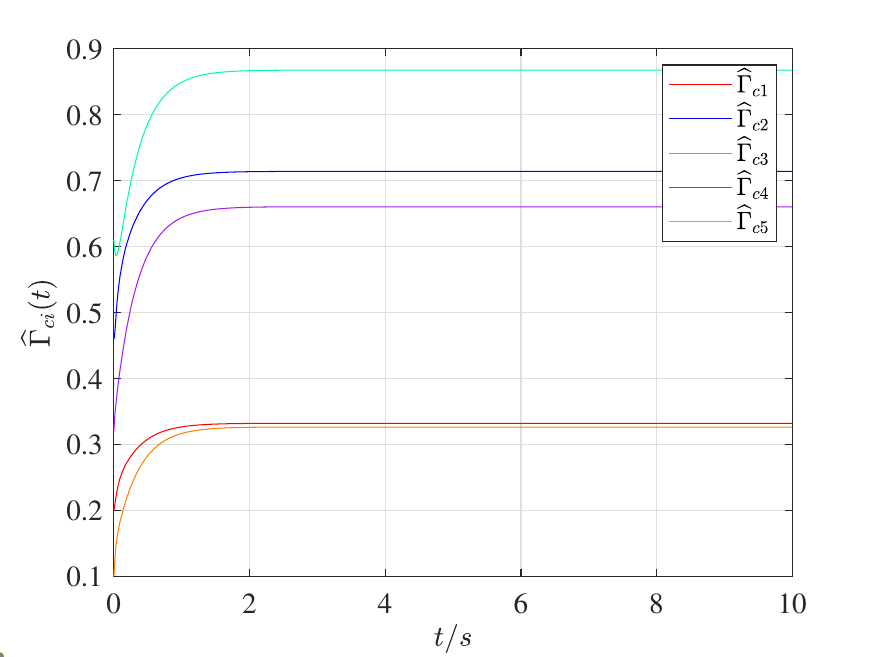}
	\caption{Weight convergence of the critic networks.}
	\label{fig.9}
\end{figure}
\begin{figure}[htbp]
	\centering
	\includegraphics[width=0.9\linewidth]{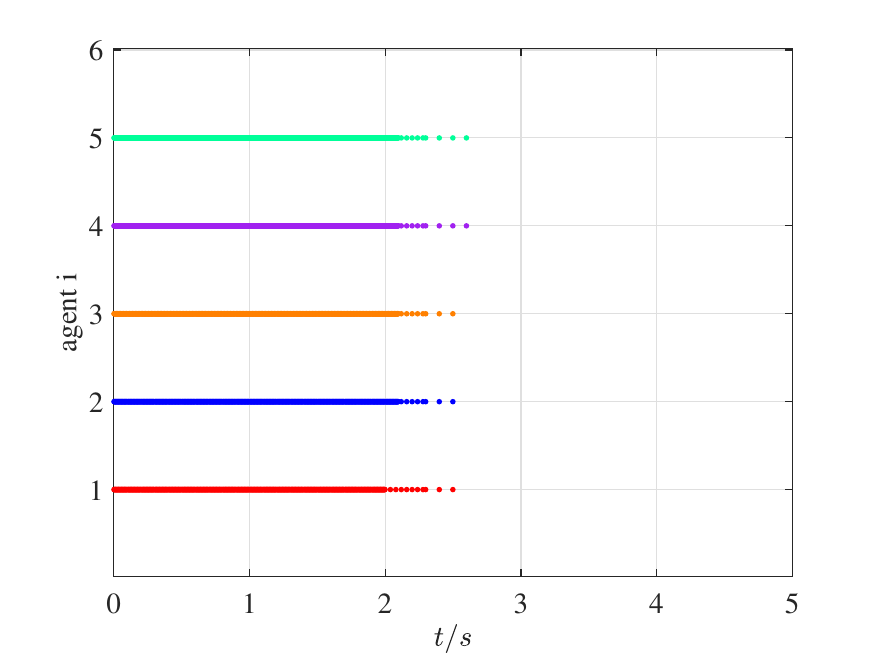}
	\caption{Trigger time of each agent.}
	\label{fig.10}
\end{figure}

The simulation results are depicted in Figures \ref{fig.3} - \ref{fig.10}. Figure \ref{fig.3} illustrates the trajectories of each agent and the formation process. It is evident from these figures that all followers are able to maintain a certain formation and follow the leader agent at the same speed. Figures \ref{fig.4} and \ref{fig.5} depict the formation errors in the horizontal and vertical directions for the MAS. From the figures, it can be observed that the formation errors of the MAS converge asymptotically to zero after 2 seconds, confirming the correctness of the theory. Figures \ref{fig.6} and \ref{fig.7} represent the control inputs in the horizontal and vertical directions for the MAS. From the figures, it is evident that the designed optimal control law can achieve the desired results with minimal consumption of control resources. Additionally, when the system is under a DoS attack, the communication channels between agents are interrupted, resulting in control signals becoming zero at that time. 
Figure \ref{fig.8} displays the cumulative cost for all follower agents under the event-triggered ADP policy. Figure \ref{fig.9} illustrates the pattern of weight changes in the neural network. It can be observed that the weight update pattern is non-periodic and converges to a constant, demonstrating that the weight estimation error is globally uniformly ultimately bounded. 
Figure \ref{fig.10} illustrates the event-triggering instants for each agent, demonstrating that by designing the event-triggering mechanism, the controller signals only need to be updated at the event-triggering instants, reducing the consumption of control resources.

\section{Conclusion}\label{sec_6}
For a class of nonlinear MASs under DoS attacks, this paper proposes an event-triggered optimal formation control scheme to address the formation tracking problem of MASs. An ADP algorithm is employed to design a single-network structure based on critic neural networks to approximate the optimal control policy. Achieving the intended control objectives by adapting the control gain of the optimal control algorithm. Additionally, we design an event-triggered mechanism so that the updating of optimal control policies and neural network weights occurs only when the triggering conditions are met, thereby reducing the consumption of communication resources. The Lyapunov stability theory is utilized to prove that the formation tracking error of the MAS exponentially converges to zero. Simulation results demonstrate that when the length rate and frequency of DoS attacks are both below the given upper bounds, the designed controller can achieve the formation control objective, validating the effectiveness of the control algorithm.

\bibliographystyle{elsarticle-num}
\bibliography{wileyNJD-AMA}

\begin{thebibliography}{10}
\expandafter\ifx\csname url\endcsname\relax
  \def\url#1{\texttt{#1}}\fi
\expandafter\ifx\csname urlprefix\endcsname\relax\def\urlprefix{URL }\fi
\expandafter\ifx\csname href\endcsname\relax
  \def\href#1#2{#2} \def\path#1{#1}\fi

\bibitem{mcarthur2004design}
S.~D. McArthur, S.~M. Strachan, G.~Jahn, The design of a multi-agent
  transformer condition monitoring system, IEEE Transactions on Power Systems
  19 (2004) 1845--1852.

\bibitem{han2013cooperative}
J.~Han, C.-h. Wang, G.-x. Yi, Cooperative control of uav based on multi-agent
  system, in: 2013 IEEE 8th Conference on Industrial Electronics and
  Applications (ICIEA), IEEE, 2013, pp. 96--101.

\bibitem{chen2020multi}
Y.-L. Chen, X.-W. Ma, G.-Q. Bai, Y.~Sha, J.~Liu, Multi-autonomous underwater
  vehicle formation control and cluster search using a fusion control strategy
  at complex underwater environment, Ocean Engineering 216 (2020) 108048.

\bibitem{ren2008distributed}
W.~Ren, N.~Sorensen, Distributed coordination architecture for multi-robot
  formation control, Robotics and Autonomous Systems 56~(4) (2008) 324--333.

\bibitem{zhang2023finite}
Z.~Zhang, K.~Yang, L.~Ouyang, Finite-time adrc formation control for uncertain
  nonaffine nonlinear multi-agent systems with prescribed performance and input
  saturation, Robotica 41~(10) (2023) 3079--3100.

\bibitem{li2021fuzzy}
Y.~Li, J.~Zhang, S.~Tong, Fuzzy adaptive optimized leader-following formation
  control for second-order stochastic multiagent systems, IEEE Transactions on
  Industrial Informatics 18~(9) (2021) 6026--6037.

\bibitem{fei2020neural}
Y.~Fei, P.~Shi, C.-C. Lim, Neural network adaptive dynamic sliding mode
  formation control of multi-agent systems, International Journal of Systems
  Science 51~(11) (2020) 2025--2040.

\bibitem{zhang2014leader}
H.~Zhang, J.~Zhang, G.-H. Yang, Y.~Luo, Leader-based optimal coordination
  control for the consensus problem of multiagent differential games via fuzzy
  adaptive dynamic programming, IEEE Transactions on Fuzzy Systems 23~(1)
  (2014) 152--163.

\bibitem{cai2020fuzzy}
Y.~Cai, H.~Zhang, K.~Zhang, C.~Liu, Fuzzy adaptive dynamic programming-based
  optimal leader-following consensus for heterogeneous nonlinear multi-agent
  systems, Neural Computing and Applications 32~(13) (2020) 8763--8781.

\bibitem{zhao2023fuzzy}
H.~Zhao, H.~Wang, N.~Xu, X.~Zhao, S.~Sharaf, Fuzzy approximation-based optimal
  consensus control for nonlinear multiagent systems via adaptive dynamic
  programming, Neurocomputing 553 (2023) 126529.

\bibitem{tang2023adaptive}
F.~Tang, H.~Wang, L.~Zhang, N.~Xu, A.~M. Ahmad, Adaptive optimized consensus
  control for a class of nonlinear multi-agent systems with asymmetric input
  saturation constraints and hybrid faults, Communications in Nonlinear Science
  and Numerical Simulation 126 (2023) 107446.

\bibitem{zhang2017distributed}
J.~Zhang, H.~Zhang, T.~Feng, Distributed optimal consensus control for
  nonlinear multiagent system with unknown dynamic, IEEE transactions on neural
  networks and learning systems 29~(8) (2017) 3339--3348.

\bibitem{amullen2016model}
E.~M. Amullen, S.~Shetty, L.~H. Keel, Model-based resilient control for a
  multi-agent system against denial of service attacks, in: 2016 World
  Automation Congress (WAC), IEEE, 2016, pp. 1--6.

\bibitem{yang2020observer}
H.~Yang, D.~Ye, Observer-based fixed-time secure tracking consensus for
  networked high-order multiagent systems against dos attacks, IEEE
  Transactions on Cybernetics 52~(4) (2020) 2018--2031.

\bibitem{amullen2016secured}
E.~M. Amullen, S.~Shetty, L.~H. Keel, Secured formation control for multi-agent
  systems under dos attacks, in: 2016 IEEE Symposium on Technologies for
  Homeland Security (HST), IEEE, 2016, pp. 1--6.

\bibitem{li2022fully}
W.~Li, H.~Zhang, W.~Wang, Z.~Cao, Fully distributed event-triggered
  time-varying formation control of multi-agent systems subject to
  mode-switching denial-of-service attacks, Applied Mathematics and Computation
  414 (2022) 126645.

\bibitem{dong2020leader}
T.~Dong, Y.~Gong, Leader-following secure consensus for second-order
  multi-agent systems with nonlinear dynamics and event-triggered control
  strategy under dos attack, Neurocomputing 416 (2020) 95--102.

\bibitem{tian2022event}
Y.~Tian, S.~Tian, H.~Li, Q.~Han, X.~Wang, Event-triggered security consensus
  for multi-agent systems with markov switching topologies under dos attacks,
  Energies 15~(15) (2022) 5353.

\bibitem{li2024bipartite}
T.~Li, S.~Li, Y.~Wang, Y.~Hui, J.~Han, Bipartite formation control of nonlinear
  multi-agent systems with fixed and switching topologies under aperiodic dos
  attacks, Electronics 13~(4) (2024) 696.

\bibitem{pham2023adaptive}
T.~V. Pham, Q.~T.~T. Nguyen, Adaptive formation control of nonlinear
  multi-agent systems with dynamic event-triggered communication, Systems \&
  Control Letters 181 (2023) 105652.

\bibitem{luo2023event}
Y.~Luo, X.~Gao, A.~Li, A.~Kashkynbayev, Event-triggered finite-time formation
  control for second-order leader-following multi-agent systems with nonlinear
  term time delay, International Journal of Control 96~(10) (2023) 2636--2650.

\bibitem{cai2021fixed}
Y.~Cai, H.~Zhang, Y.~Wang, J.~Zhang, Q.~He, Fixed-time time-varying formation
  tracking for nonlinear multi-agent systems under event-triggered mechanism,
  Information Sciences 564 (2021) 45--70.

\bibitem{wang2022event2}
X.~Wang, Z.~Quan, Y.~Li, Y.~Liu, Event-triggered trajectory-tracking guidance
  for reusable launch vehicle based on neural adaptive dynamic programming,
  Neural Computing and Applications 34~(21) (2022) 18725--18740.

\bibitem{zhao2019event}
W.~Zhao, W.~Yu, H.~Zhang, Event-triggered optimal consensus tracking control
  for multi-agent systems with unknown internal states and disturbances,
  Nonlinear Analysis: Hybrid Systems 33 (2019) 227--248.

\bibitem{zhao2019distributed}
W.~Zhao, H.~Zhang, Distributed optimal coordination control for nonlinear
  multi-agent systems using event-triggered adaptive dynamic programming
  method, ISA transactions 91 (2019) 184--195.

\bibitem{dou2022event}
L.~Dou, S.~Cai, X.~Zhang, X.~Su, R.~Zhang, Event-triggered-based adaptive
  dynamic programming for distributed formation control of multi-uav, Journal
  of the Franklin Institute 359~(8) (2022) 3671--3691.

\bibitem{wang2022event}
J.~Wang, Z.~Zhang, B.~Tian, Q.~Zong, Event-based robust optimal consensus
  control for nonlinear multiagent system with local adaptive dynamic
  programming, IEEE Transactions on Neural Networks and Learning Systems 35~(1)
  (2022) 1073--1086.

\bibitem{liu2023aperiodically}
L.~Liu, J.~Cao, F.~E. Alsaadi, Aperiodically intermittent event-triggered
  optimal average consensus for nonlinear multi-agent systems, IEEE
  Transactions on Neural Networks and Learning Systems (2023) 1--15.

\bibitem{yang2020online}
Y.~Yang, D.-W. Ding, H.~Xiong, Y.~Yin, D.~C. Wunsch, Online
  barrier-actor-critic learning for h$\infty$ control with full-state
  constraints and input saturation, Journal of the Franklin Institute 357~(6)
  (2020) 3316--3344.

\bibitem{vamvoudakis2014event}
K.~G. Vamvoudakis, Event-triggered optimal adaptive control algorithm for
  continuous-time nonlinear systems, IEEE/CAA Journal of Automatica Sinica
  1~(3) (2014) 282--293.

\end{thebibliography}

\end{document}